%% file: main.tex
\documentclass[twocolumn,twocolappendix]{aastex631}
\usepackage{CJK}
\usepackage[caption=false]{subfig}

\received{2022 March 15}
\revised{2022 May 30}
\accepted{2022 May 30}
\submitjournal{\textit{The Astrophysical Journal}}

\shorttitle{SN~2021yja}
\shortauthors{Hosseinzadeh et al.}

\newcommand\EBVMW{$E(B-V)_\mathrm{MW} = 0.0152_{-0.0026}^{+0.0031}$~mag}
\newcommand\EBVMWdust{$E(B-V)_\mathrm{MW} = 0.0191$~mag}
\newcommand\EBVhost{$E(B-V)_\mathrm{host} = 0.085_{-0.014}^{+0.017}$~mag}
\newcommand\EWDIB{$W_\lambda = 0.0050_{-0.0020}^{+0.0033}$~nm}
\newcommand\EBVtot{$E(B-V)=0.104$~mag}

\newcommand\radius{$2030^{+260}_{-230}\ R_\sun$}
\newcommand\explosion{MJD $59464.40 \pm 0.06$}
\newcommand\deltaexplosion{$5.4 \pm 1.4$ rest-frame hours}
\newcommand\tmax{MJD 59496.3}
\newcommand\tmaxphase{31.7 days}

\begin{document}

\title{Weak Mass Loss from the Red Supergiant Progenitor of the Type~II SN~2021yja}

\correspondingauthor{Griffin Hosseinzadeh}
\email{griffin0@arizona.edu}

\input{affil.tex}

\author[0000-0002-0832-2974]{Griffin Hosseinzadeh}
\UA
\author[0000-0002-5740-7747]{Charles D.\ Kilpatrick}
\CIERA
\author[0000-0002-7937-6371]{Yize Dong \begin{CJK*}{UTF8}{gbsn}(董一泽)\end{CJK*}}
\UCD
\author[0000-0003-4102-380X]{David J.\ Sand}
\UA

\author[0000-0003-0123-0062]{Jennifer E.\ Andrews}
\GeminiNorth
\author[0000-0002-4924-444X]{K.\ Azalee Bostroem}
\DiRAC\UW
\author[0000-0003-0549-3281]{Daryl Janzen}
\USask
\author[0000-0001-5754-4007]{Jacob E.\ Jencson}
\UA
\author[0000-0001-9589-3793]{Michael Lundquist}
\Keck
\author[0000-0002-7015-3446]{Nicolas E.\ Meza Retamal}
\UCD
\author[0000-0002-0744-0047]{Jeniveve Pearson}
\UA
\author[0000-0001-8818-0795]{Stefano Valenti}
\UCD
\author[0000-0003-2732-4956]{Samuel Wyatt}
\UA

\author[0000-0003-0035-6659]{Jamison Burke}
\LCO\UCSB
\author[0000-0002-1125-9187]{Daichi Hiramatsu}
\CfA\IAIFI
\author[0000-0003-4253-656X]{D.\ Andrew Howell}
\LCO\UCSB
\author[0000-0001-5807-7893]{Curtis McCully}
\LCO\UCSB
\author{Megan Newsome}
\LCO\UCSB
\author[0000-0003-0209-9246]{Estefania Padilla Gonzalez}
\LCO\UCSB
\author[0000-0002-7472-1279]{Craig Pellegrino}
\LCO\UCSB
\author[0000-0003-0794-5982]{Giacomo Terreran}
\LCO\UCSB

\author[0000-0002-4449-9152]{Katie Auchettl}
\Melbourne\ASTROthreeD\UCSC
\author[0000-0002-5680-4660]{Kyle W.\ Davis}
\UCSC
\author[0000-0002-2445-5275]{Ryan J. Foley}
\UCSC
\author[0000-0003-2736-5977]{Hao-Yu Miao \begin{CJK*}{UTF8}{bsmi}(繆皓宇)\end{CJK*}}
\NCU
\author[0000-0001-8415-6720]{Yen-Chen Pan \begin{CJK*}{UTF8}{bsmi}(潘彥丞)\end{CJK*}}
\NCU
\author[0000-0002-4410-5387]{Armin Rest}
\STScI\JHU
\author[0000-0003-2445-3891]{Matthew R.\ Siebert}
\UCSC
\author[0000-0002-5748-4558]{Kirsty Taggart}
\UCSC
\author[0000-0002-4283-5159]{Brad E.\ Tucker}
\Stromlo\NCPAS\ASTROthreeD

\author[0000-0002-6604-8838]{Feng Lin Cyrus Leung}
\Thacher
\author[0000-0002-9486-818X]{Jonathan J.\ Swift}
\Thacher
\author[0000-0001-7823-2627]{Grace Yang}
\Thacher

\author[0000-0003-0227-3451]{Joseph P.\ Anderson}
\ESO
\author[0000-0002-5221-7557]{Chris Ashall}
\IfA
\author[0000-0002-3256-0016]{Stefano Benetti}
\OAPD
\author[0000-0001-6272-5507]{Peter J.\ Brown}
\TAMU\Mitchell
\author[0000-0003-4553-4033]{R\'egis Cartier}
\GeminiSouth
\author[0000-0002-1066-6098]{Ting-Wan Chen \begin{CJK*}{UTF8}{bsmi}(陳婷琬)\end{CJK*}}
\OKC
\author[0000-0003-3142-5020]{Massimo Della Valle}
\Capodimonte\INFNNapoli\ICRANet
\author[0000-0002-1296-6887]{Llu\'is Galbany}
\ICE\IEEC
\author[0000-0001-6395-6702]{Sebastian Gomez}
\STScI
\author[0000-0002-1650-1518]{Mariusz Gromadzki}
\Warsaw
\author[0000-0002-6703-805X]{Joshua Haislip}
\UNC
\author[0000-0003-1039-2928]{Eric Y.\ Hsiao}
\FSU
\author[0000-0002-3968-4409]{Cosimo Inserra}
\Cardiff
\author[0000-0001-8738-6011]{Saurabh W.\ Jha}
\Rutgers
\author[0000-0002-0440-9597]{Thomas L.\ Killestein}
\Warwick
\author[0000-0003-3642-5484]{Vladimir Kouprianov}
\UNC
\author[0000-0001-9598-8821]{Alexandra Kozyreva}
\MPIA
\author[0000-0003-3939-7167]{Tom\'as E.\ M\"uller-Bravo}
\ICE
\author[0000-0002-2555-3192]{Matt~Nicholl}
\Birmingham
\author[0000-0003-2814-4383]{Emmy Paraskeva}
\IAASARS\NKUA
\author[0000-0002-5060-3673]{Daniel E.\ Reichart}
\UNC
\author[0000-0003-4501-8100]{Stuart Ryder}
\Macquarie\AAARC
\author[0000-0002-9301-5302]{Melissa Shahbandeh}
\FSU
\author[0000-0003-4631-1149]{Ben Shappee}
\IfA
\author[0000-0001-5510-2424]{Nathan Smith}
\UA
\author[0000-0002-1229-2499]{David R.\ Young}
\QUB

\begin{abstract}

We present high-cadence optical, ultraviolet (UV), and near-infrared data of the nearby ($D\approx23$~Mpc) Type~II supernova (SN) 2021yja. Many Type II SNe show signs of interaction with circumstellar material (CSM) during the first few days after explosion, implying that their red supergiant (RSG) progenitors experience episodic or eruptive mass loss. However, because it is difficult to discover SNe early, the diversity of CSM configurations in RSGs has not been fully mapped. SN\,2021yja, first detected within ${\approx}5.4$ hours of explosion, shows some signatures of CSM interaction (high UV luminosity, radio and x-ray emission) but without the narrow emission lines or early light-curve peak that can accompany CSM. Here we analyze the densely sampled early light curve and spectral series of this nearby SN to infer the properties of its progenitor and CSM. We find that the most likely progenitor was an RSG with an extended envelope, encompassed by low-density CSM. We also present archival Hubble Space Telescope imaging of the host galaxy of SN~2021yja, which allows us to place a stringent upper limit of ${\lesssim}9\ M_\sun$ on the progenitor mass. However, this is in tension with some aspects of the SN evolution, which point to a more massive progenitor. Our analysis highlights the need to consider progenitor structure when making inferences about CSM properties, and that a comprehensive view of CSM tracers should be made to give a fuller view of the last years of RSG evolution.

\end{abstract}

\keywords{Circumstellar matter (241), Core-collapse supernovae (304), Stellar mass loss (1613), Supernovae (1668), Type II supernovae (1731)}

\section{Introduction} \label{sec:intro}
Mass loss in massive stars ($M_\mathrm{ZAMS} \gtrsim 8\,M_\sun$) is among the most poorly understood aspects of stellar evolution \citep{smith_mass_2014}, and core-collapse supernovae (CCSNe) provide a complementary window into studying mass-loss processes that the stars of the Milky Way and its satellites cannot provide. In the case of pre-SN mass loss, the difficulty lies in the fact that mass-loss rates may change quickly, in which case the SN must be observed very shortly after explosion in order to map out the prognenitor's mass-loss history. For this reason, CCSNe in nearby galaxies---where early discovery, dense time sampling, and rich multiwavelength data sets are possible---offer a valuable opportunity for a mass-loss case study.

Type~II supernovae (SNe~II),\footnote{Throughout this work, we use ``SNe~II'' to refer to Type~IIP and Type~IIL SNe \citep{barbon_photometric_1979}, but not Type~IIb or Type~IIn SNe.} the most common CCSNe \citep{li_nearby_2011,smith_observed_2011}, are the hydrogen-rich explosions of red supergiant (RSG) stars. This has been confirmed by the direct detection of tens of SN~II progenitors in archival Hubble Space Telescope (HST) imaging \citep{smartt_observational_2015,van_dyk_supernova_2016}. Traditionally, RSGs were not thought to produce detectable amounts of CSM, but recent work has shown that CSM is nearly ubiquitous in SN~II progenitors and can affect the SN observables, particularly during the first days to weeks after explosion. For example, very early spectroscopy has revealed narrow emission lines from high-ionization states, interpreted as forming in the excited but unshocked CSM \citep{gal-yam_wolf-rayet-like_2014,smith_ptf11iqb:_2015,khazov_flash_2016,yaron_confined_2017,bullivant_sn_2018,hosseinzadeh_short-lived_2018,soumagnac_sn_2020,bruch_large_2021,hiramatsu_electron-capture_2021,terreran_early_2022}. \cite{khazov_flash_2016} and \cite{bruch_large_2021} have observed these in a large fraction of SNe~II. Likewise, \cite{morozova_unifying_2017,morozova_measuring_2018} show through radiation-hydrodynamical modeling that early SN~II light curves require a CSM component to reproduce the fast rise and sometimes even an early peak \citep[see also][]{gonzalez-gaitan_rise-time_2015,forster_delay_2018}.

Given that both of these CSM signatures disappear shortly after explosion, early discovery, classification, and follow-up are critical to understanding circumstellar interaction in CCSNe, and therefore to understanding mass loss in CCSN progenitors. Specialized surveys like the Distance Less Than 40\,Mpc (DLT40) Survey \citep{tartaglia_early_2018} observe nearby galaxies frequently (twice per day for DLT40 in the southern hemisphere) with the goal of discovering new transients shortly after explosion, when they are still very faint, and announce discoveries immediately so that spectroscopic classification can occur shortly after. Such projects, used in concert with wide-field transient searches such as the Asteroid Terrestrial-impact Last Alert System \citep[ATLAS;][]{tonry_atlas_2018}, the Young Supernova Experiment \citep[YSE;][]{jones_young_2021}, the All-Sky Automated Survey for Supernovae \citep{shappee_man_2014}, and the Zwicky Transient Facility \citep{bellm_zwicky_2019}, can place strong limits on the explosion epochs of nearby SNe and allow for the most comprehensive early data sets to be acquired. Likewise, rapid robotic follow-up with facilities like Las Cumbres Observatory \citep{brown_cumbres_2013} and the Neil Gehrels Swift Observatory \citep{gehrels_swift_2004} allow us to trace the SN ejecta in detail as it collides with and overruns the material ejected in the years before explosion.

Here we consider the case of SN~2021yja, a nearby ($D\approx23$~Mpc) SN~II observed within hours of explosion and followed extensively in the optical and ultraviolet (UV) for the first 150~days of its evolution. The site of SN~2021yja was previously observed with HST to a depth that allows us to place a very strong constraint on the progenitor luminosity. In \S\ref{sec:obs}, we describe the discovery and observations of SN~2021yja and derive the properties of its host galaxy required for our analysis. In \S\ref{sec:analysis}, we analyze the light curves, spectra, and pre-explosion imaging of SN~2021yja in order to constrain its progenitor properties, including the presence of any CSM, and we compare it to other well-observed SNe~II in the literature. Finally, in \S\ref{sec:discuss}, we discuss the implications of our measurements for the population of SN~II progenitors and RSGs in general, with a focus on mass-loss histories.

\section{Observations and Data Reduction}\label{sec:obs}
\subsection{Discovery and Classification} \label{sec:disc}
\begin{figure*}
    \centering
    \includegraphics[width=0.27\textwidth,trim=0 -0.75in 0 0,clip]{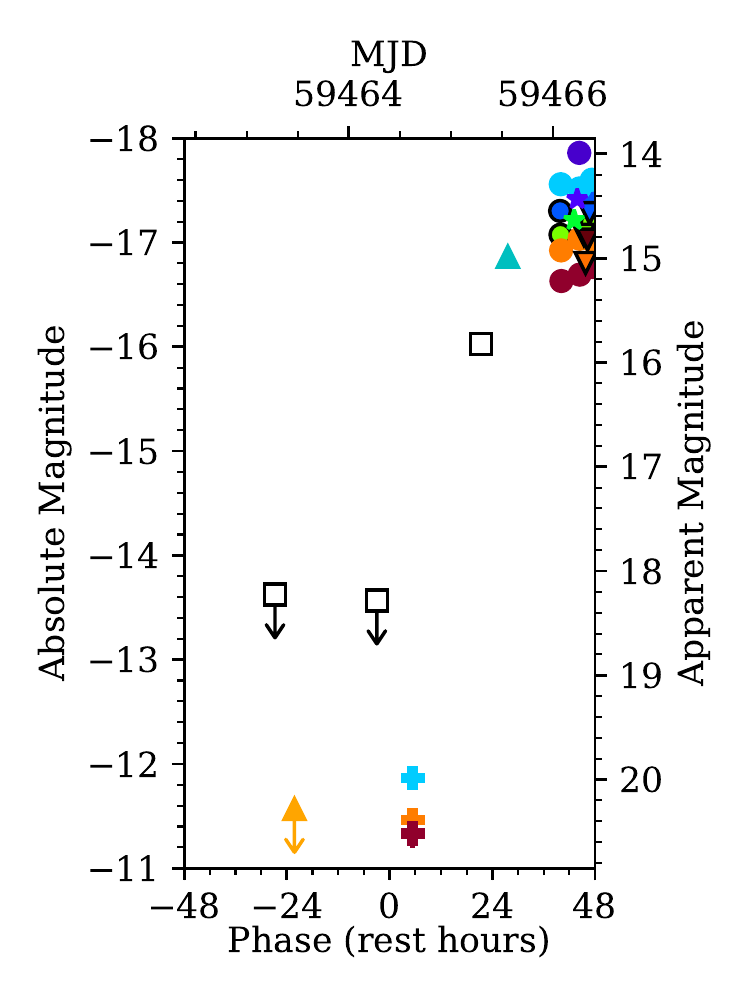}
    \includegraphics[width=0.72\textwidth]{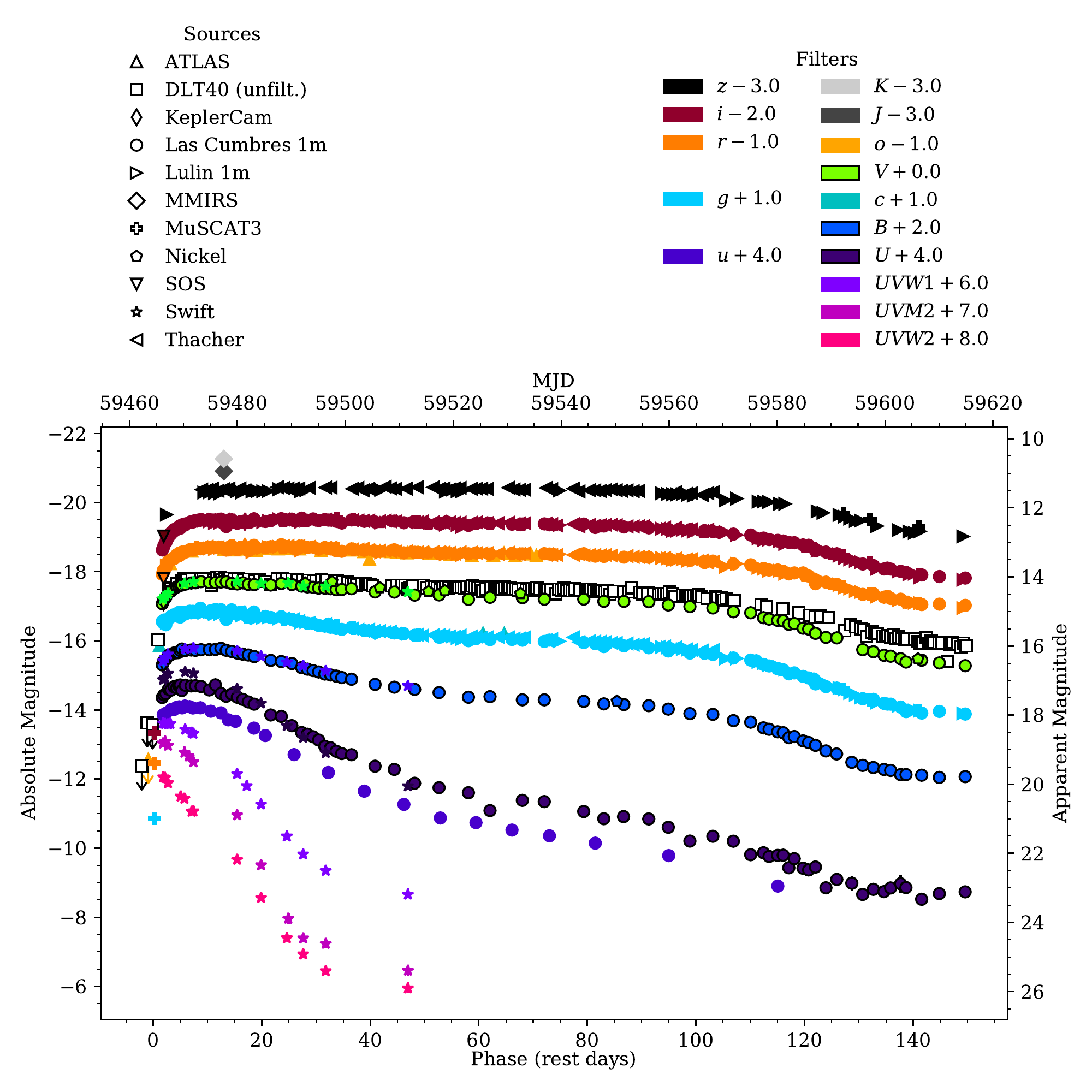}
    \caption{The main figure (right) shows the multiband UV-optical-infrared light curve of SN~2021yja through the end of its plateau. Phase is given in rest-frame days after estimated explosion. Note the unusually long (${\approx}140$~days) plateau and slow fall onto the radioactive-decay-powered tail (the last 3 epochs). The inset (left) shows the light curve within 48 hours of explosion, with no filter offsets. In particular, note the MuSCAT3 data (plus signs), which constrains the explosion epoch even more tightly than the nondetections by ATLAS and DLT40 (see \S~\ref{sec:sc}). (The data used to create this figure are available.)}
    \label{fig:phot}
\end{figure*}

SN~2021yja was discovered by ATLAS on 2021-09-08.55 (all dates are in UTC) at $c = 15.334 \pm 0.007$~mag and was not detected on 2021-09-06.48 to a $3\sigma$ limit of $o > 19.22$~mag \citep{smith_atlas21bidw_2021,tonry_atlas_2021}. Its J2000 coordinates, as measured by Gaia Photometric Science Alerts, are $\alpha = 03\textsuperscript{h}24\textsuperscript{m}21\fs180$, $\delta = -21\degr33'56\farcs20$ \citep{hodgkin_gaiaalerts_2021}, $99''$ southwest of the center of its host galaxy, NGC~1325.\footnote{NGC~1325 previously hosted SN~1975S, which was not spectroscopically classified \citep{wegner_photometric_1977}.} It was recovered in earlier unfiltered (357--871~nm) imaging by the DLT40 Survey on 2021-09-08.29 at $16.431 \pm 0.028$~mag and was not detected on 2021-09-07.28 to a $3\sigma$ limit of 18.671~mag. \cite{kilpatrick_at2021yja:_2021} also recovered it at $g=20.4 \pm 0.2$~mag and $r = i = 20.7 \pm 0.2$~mag in publicly available imaging taken on 2021-09-07.63 with the Multicolor Simultaneous Camera for Studying Atmospheres of Transiting Exoplanets 3 (MuSCAT3) on Las Cumbres Observatory's 2\,m Faulkes Telescope North \citep{narita_muscat3:_2020}, taken as part of the Faulkes Telescope Project's education and public outreach program.\footnote{After a final reduction of these images, we report updated magnitudes in Figure~\ref{fig:phot} and its associated online-only table.} The very low luminosities implied by these magnitudes meant that it was not clear in real time whether these MuSCAT3 images were taken before or after explosion (see \S\ref{sec:sc}).

SN~2021yja was initially classified as a young SN~II by \cite{pellegrino_global_2021} using a spectrum taken by the Global Supernova Project (GSP) on 2021-09-09.51, 15 hours after the ATLAS discovery report. Their classification was based on spectroscopic comparisons with other SNe~II using Superfit \citep{howell_gemini_2005} and the Supernova Identification code \citep[SNID;][]{blondin_determining_2007}. Shortly afterward, it was reclassified as a young stripped-envelope SN by \cite{williamson_transient_2021} using a second spectrum taken by the GSP on 2021-09-09.60. They argued that the spectrum also resembled other young SN~Ic spectra, and that the putative high-velocity H$\alpha$ line could have been misidentified silicon 635.5~nm. Several days later, after stronger hydrogen lines had developed, \cite{deckers_epessto+_2021,deckers_epessto+_2021a} restored the SN~II classification using a spectrum taken on 2021-09-14.31 by the advanced extended Public ESO Spectroscopic Survey of Transient Objects \citep[ePESSTO+;][]{smartt_pessto:_2015}.

\subsection{Follow-up Photometry and Spectroscopy}
We obtained UV, optical, and infrared photometry of SN~2021yja using the Sinistro cameras on Las Cumbres Observatory's network of 1\,m telescopes \citep{brown_cumbres_2013},\footnote{Sinistro data were obtained under both the GSP and the YSE programs.} MuSCAT3 on Las Cumbres Observatory's 2\,m Faulkes Telescope North \citep{narita_muscat3:_2020}, an Andor iKON-L 936 BV camera on the 0.7\,m telescope at Thacher Observatory \citep{swift_renovated_2022}, KeplerCam on the 1.2\,m telescope at F.~L.~Whipple Observatory \citep{szentgyorgyi_keplercam_2005}, the Lulin Compact Imager on the 1\,m telescope at Lulin Observatory, the MMT and Magellan Infrared Spectrograph (MMIRS) on the MMT \citep{mcleod_mmt_2012}, the Direct Imaging Camera on the Nickel Telescope \citep{stone_ccd_1990}, the Ultraviolet/Optical Telescope on Swift \citep{roming_swift_2005}, and Andor Apogee Alta F47 cameras on the Panchromatic Robotic Optical Monitoring and Polarimetry Telescopes at Cerro Tololo Inter-American Observatory and Meckering Observatory \citep{reichart_prompt:_2005} as part of the DLT40 Survey. See Appendix~\ref{sec:photred} for details of the photometry reduction. We also include publicly available photometry from the ATLAS forced photometry server \citep{tonry_atlas:_2018,smith_design_2020} and the Supernova Observations and Simulations group \citep{martinez_follow-up_2021}.
Figure~\ref{fig:phot} shows the light curve of SN~2021yja; the data are available in machine-readable format in the online journal. 

In addition, we obtained high-resolution follow-up imaging with the Gemini South Adaptive Optics Imager (GSAOI) at Cerro Pach\'on, Chile \citep{mcgregor_gemini_2004}.  We observed the site of SN\,2021yja on 2021-12-18 in $H$-band using an on-off pattern for 300~s of total on-source exposure.  Following procedures in \citet{kilpatrick_cool_2021}, we performed image calibration using dark and flat-field frames obtained in the instrumental configuration.  

\begin{figure*}[p]
    \centering
    \includegraphics[width=\textwidth]{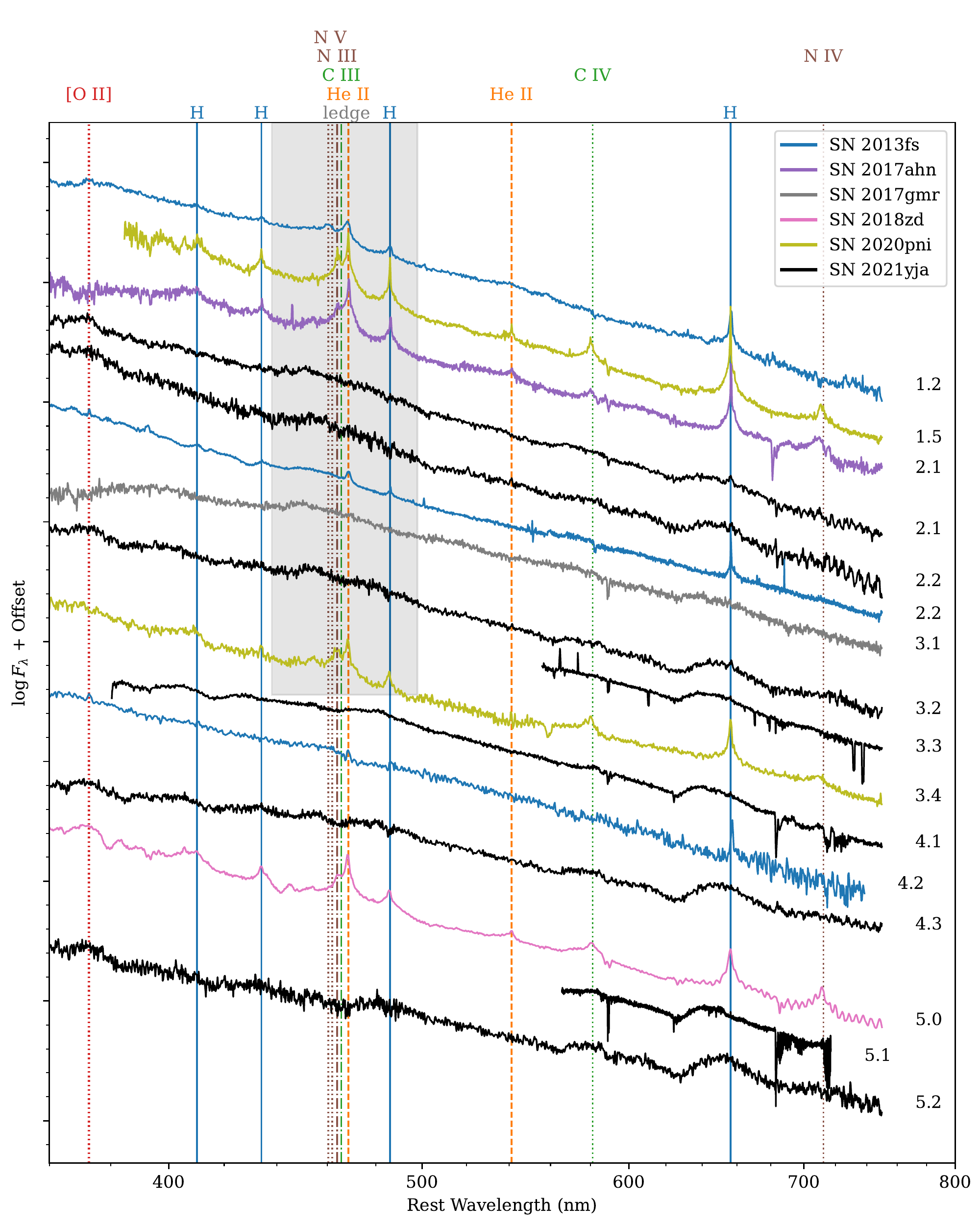}
    \caption{The early spectroscopic evolution of SN~2021yja compared to several other SNe~II with good explosion constraints. The labels to the right of each spectrum give phase in rest-frame days after explosion, with uncertainties of 0.1--0.5 days. SN~2021yja does not show strong, narrow high-ionization lines  like SNe~2013fs, 2017ahn, 2018zd, and 2020pni, an indication of short-lived circumstellar interaction. However, it does show broader features, including a mysterious broad feature around 450~nm (labeled ``ledge'') denoted by a gray bar, also seen in SN~2017gmr. (The data used to create this figure are available.)}
    \label{fig:spec_early}
\end{figure*}

After classification, we initiated a high-cadence optical spectroscopic follow-up campaign using the FLOYDS instruments on Las Cumbres Observatory's 2\,m Faulkes Telescopes North and South \citep[FTN \& FTS;][]{brown_cumbres_2013}, Binospec on the MMT \citep{fabricant_binospec:_2019}, the Wide Field Spectrograph (WiFeS) on the 2.3\,m telescope at Siding Spring Observatory \citep[SSO;][]{dopita_wide_2007,dopita_wide_2010}, the ESO Faint Object Spectrograph and Camera 2 (EFOSC2) on the
New Technology Telescope \citep[NTT;][]{buzzoni_eso_1984}; the Keck Cosmic Web Imager (KCWI) on Keck~II \citep{morrissey_keck_2018}, the Robert Stobie Spectrograph (RSS) on the
South African Large Telescope \citep[SALT;][]{smith_prime_2006}, the Kast Spectrograph on the Shane Telescope \citep{miller_kast_1994}, the Boller \& Chivens Spectrograph (B\&C) on the Bok Telescope \citep{green_steward_1995}, the High-Resolution Echelle Spectrometer (HIRES) on Keck~I \citep{vogt_hires:_1994}, the Goodman High Throughput Spectrograph on the Southern Astrophysical Research Telescope \citep[SOAR;][]{clemens_goodman_2004}, and the Dual Imaging Spectrograph (DIS) on the Astrophysical Research Consortium (ARC) 3.5\,m telescope \citep{lupton_dis:_2005}. Figure~\ref{fig:spec_early} compares the earliest spectra of SN~2021yja, taken within 6 days of estimated explosion, to other early spectra of SNe~II from \cite{yaron_confined_2017}, \cite{andrews_sn_2019}, \cite{hiramatsu_electron-capture_2021}, \cite{tartaglia_early_2021}, and \cite{terreran_early_2022}. The remaining optical spectra are plotted in Figure~\ref{fig:spec_mid}. We also obtained near-infrared spectra using the Near-Infrared Echelette Spectrometer (NIRES) on Keck~II \citep{wilson_mass_2004}, the Son of ISAAC spectrograph (SOFI) on the NTT \citep{moorwood_sofi_1998}, TripleSpec 4.1 on SOAR \citep{schlawin_design_2014}, and MMIRS on the MMT, which are plotted in Figure~\ref{fig:spec_nir}. All spectra were either observed at the parallactic angle, or an atmospheric dispersion corrector was used. Details of the spectroscopic reductions are given in Appendix~\ref{sec:specred}.

\begin{figure*}[p]
    \centering
    \includegraphics[width=\textwidth]{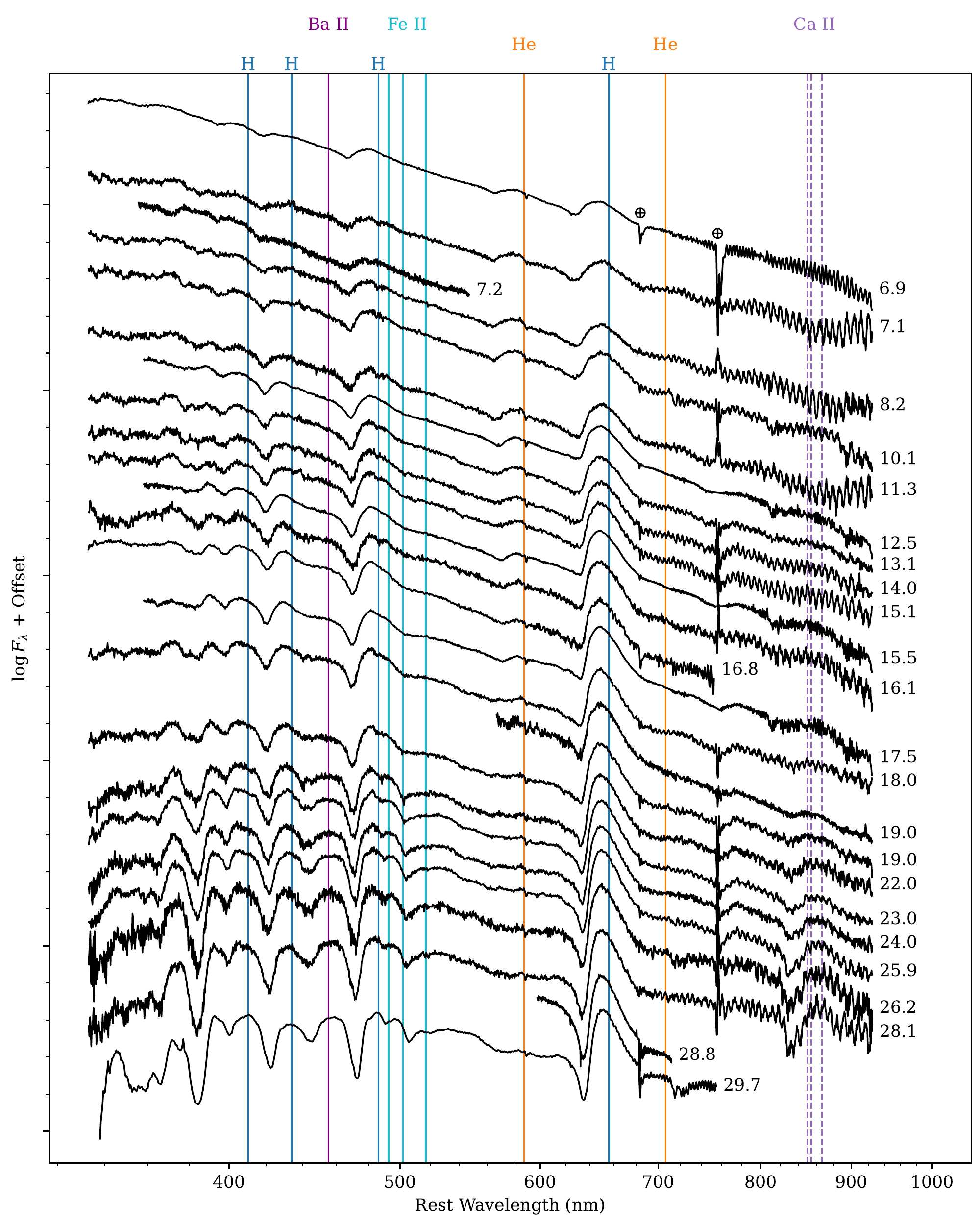}
    \caption{The spectroscopic evolution of SN~2021yja ${>}6$ days after explosion, with colored lines at the rest wavelengths of selected features. (The data used to create this figure are available.)}
    \label{fig:spec_mid}
\end{figure*}

\begin{figure*}[p]
    \centering
    \ContinuedFloat
    \includegraphics[width=\textwidth]{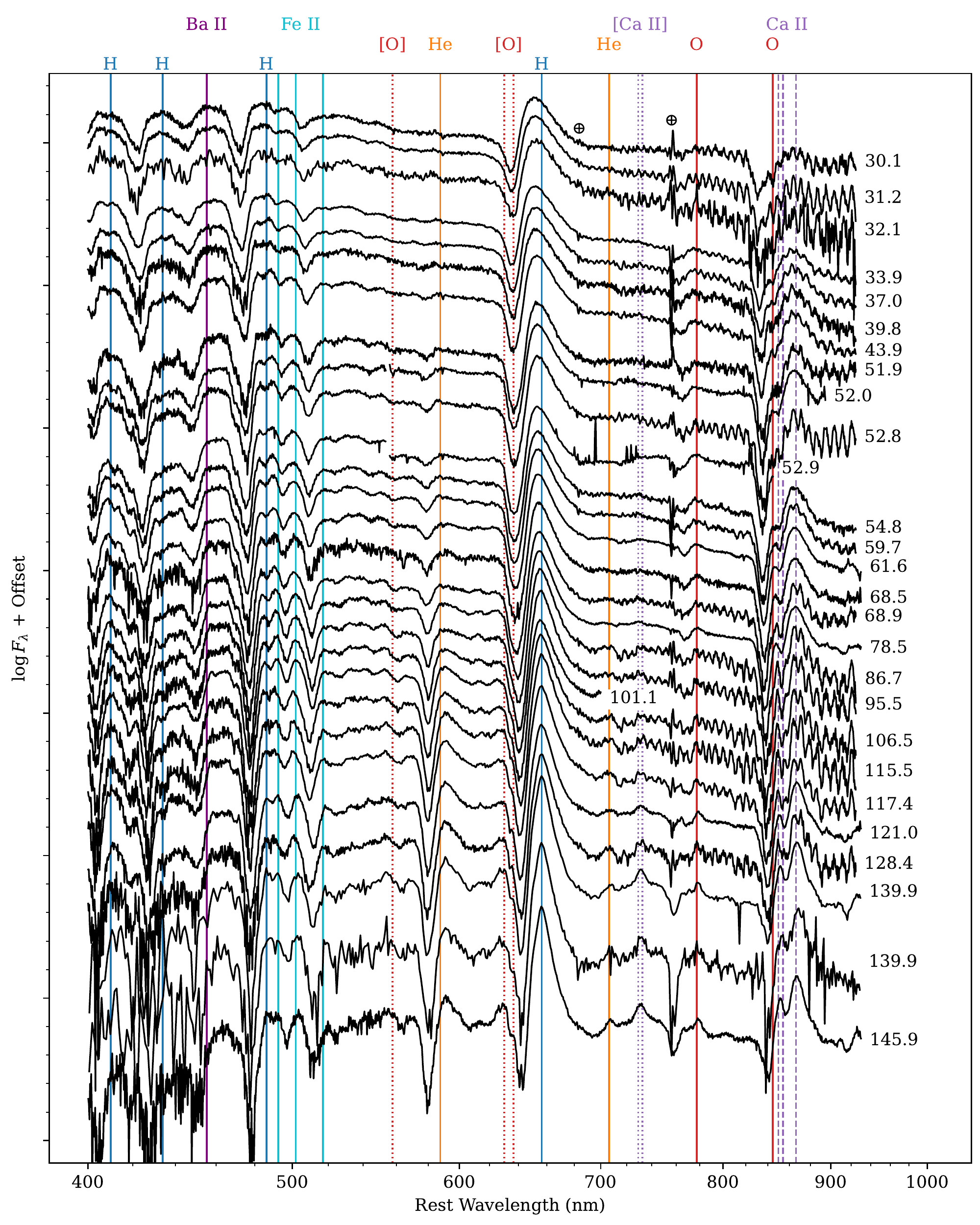}
    \caption{Continued. (The data used to create this figure are available.)}
    \label{fig:spec_late}
\end{figure*}

\begin{figure*}[p]
    \centering
    \includegraphics[width=0.94\textwidth]{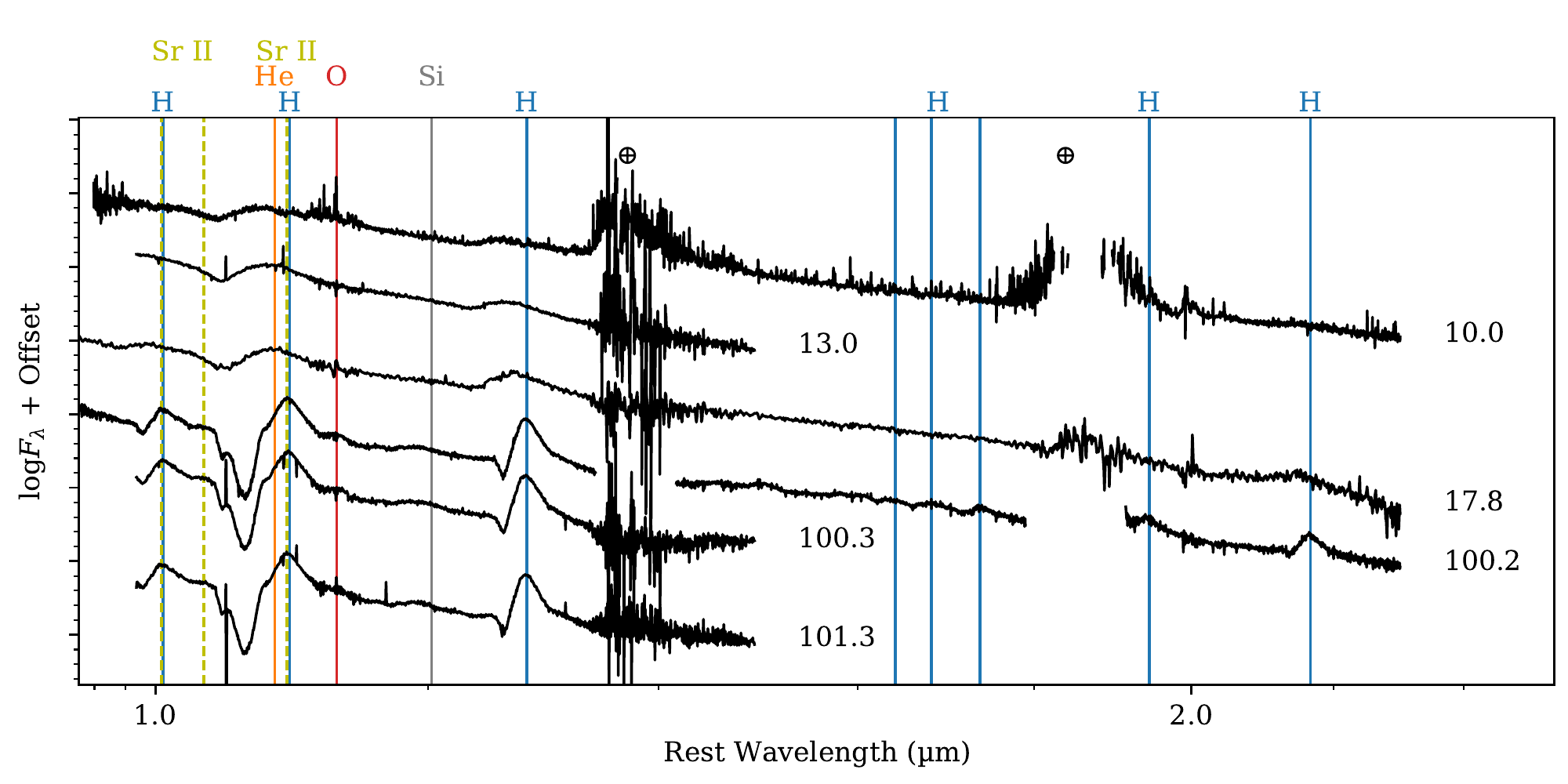}
    \caption{Near-infrared spectra of SN~2021yja. The weak helium absorption and the high-velocity helium feature in our ${>}100$~d spectra indicate that SN~2021yja belongs to the ``weak'' class of \cite{davis_carnegie_2019}. (The data used to create this figure are available.)}
    \label{fig:spec_nir}
\end{figure*}

\begin{figure*}[p]
    \includegraphics[width=0.94\textwidth]{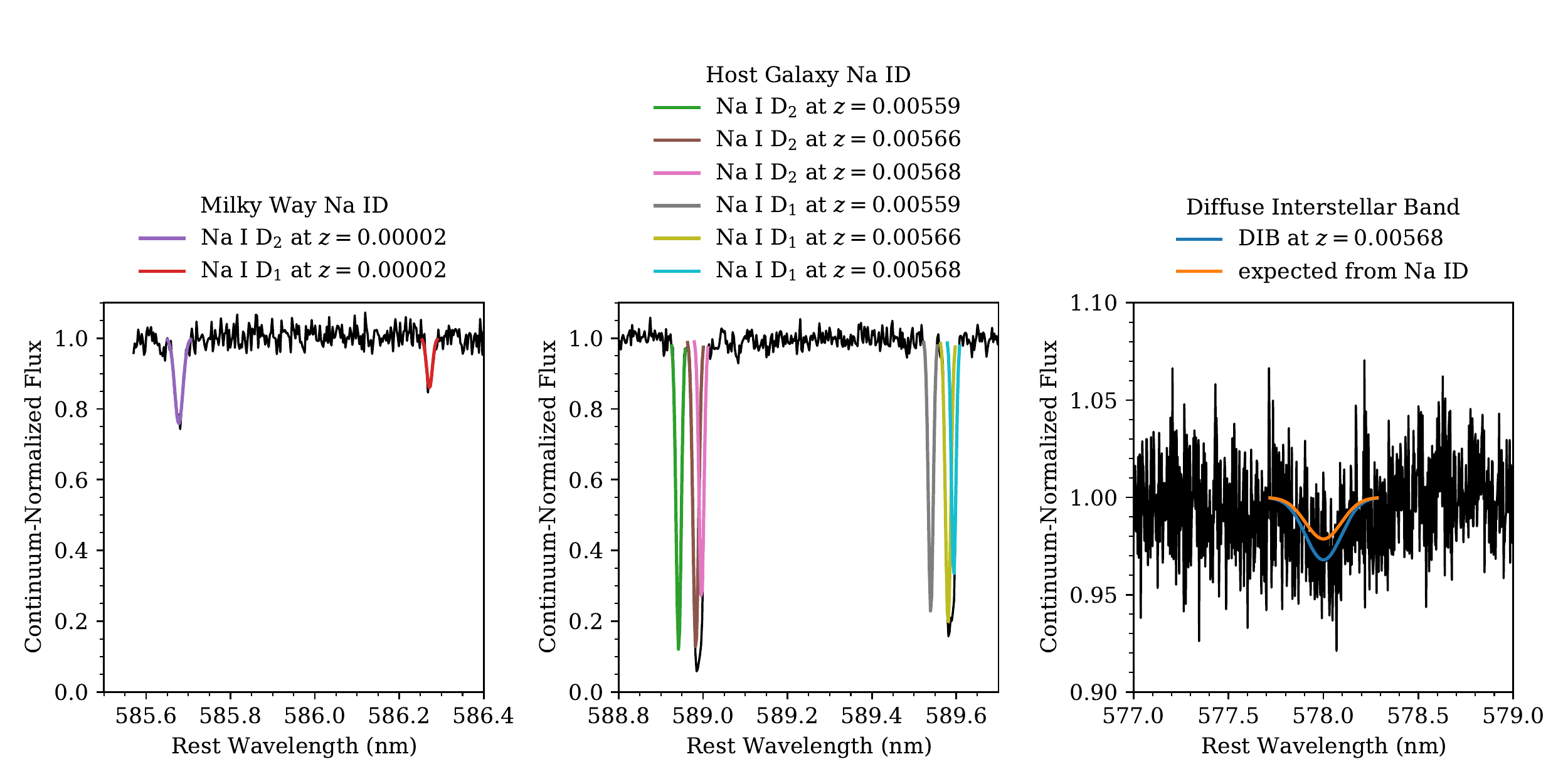}
    \caption{A high-resolution spectrum of SN~2021yja in the regions surrounding \ion{Na}{1}\,D and the diffuse interstellar band (DIB) at 578~nm. The colored lines in the left and center panels indicate four distinct line-of-sight absorbers: one in the Milky Way ($z=0.00002$) and three in the host galaxy ($z=0.00559$, 0.00566, and 0.00568). Applying the relations of \cite{poznanski_empirical_2012}, the Milky Way component is consistent with \EBVMWdust{} from the dust maps of \cite{schlafly_measuring_2011}. The strong host galaxy components indicate an additional extinction of \EBVhost{}, which is surprising given the blue observed colors of SN~2021yja. The DIB (right) confirms the significant extinction in the host galaxy. The blue Gaussian is a fit to the data, while the orange Gaussian shows the strength of DIB we would expect given the \ion{Na}{1}\,D absorption, according to the extinction relation of \cite{phillips_source_2013}. We adopt a total extinction of \EBVtot{}. (The data used to create this figure are available.)
    \label{fig:extinction}}
\end{figure*}

\subsection{Pre-explosion Imaging}

NGC~1325 and the site of SN\,2021yja were observed by HST on 1997-03-26 with the Wide Field Planetary Camera 2 (WFPC2) in F606W (PI Stiavelli, SNAP-6359).  We reduced all {\tt c0m} frames with the {\tt hst123} reduction pipeline and performed final photometry in the {\tt c0m} frames using {\tt dolphot}. In addition, we obtained DECam $grizY$ imaging taken between 2014 and 2019 and covering the site of SN\,2021yja.  We stacked and calibrated all such data using a custom pipeline based on the {\tt photpipe} imaging and photometry package \citep{rest_testing_2005}.  Finally, we downloaded pre-explosion Spitzer/IRAC imaging \citep{fazio_infrared_2004,werner_spitzer_2004,gehrz_nasa_2007} from the Spitzer Heritage Archive and combined the basic calibrated data ({\tt cbcd}) frames following methods described in \citet{kilpatrick_cool_2021}.  These pre-explosion data are analyzed in \S\ref{sec:preexplosion}.  

\subsection{Redshift}
NGC~1325 has a heliocentric redshift of $z = 0.005307 \pm 0.000005$ \citep{springob_digital_2005}. However, we observe a weak, narrow H$\alpha$ emission line in our Keck HIRES spectrum at a redshift of $z = 0.00568$, corresponding to a line-of-sight velocity of $c \Delta z = +112\mathrm{\ km\ s}^{-1}$ with respect to the galaxy core. This velocity is reasonable given the position of the SN and a typical galaxy rotation curve \citep{falcon-barroso_stellar_2017,guidi_selgifs_2018}, so we adopt this as the SN redshift.

\subsection{Extinction}\label{sec:extinction}
The equivalent widths of \ion{Na}{1}\,D absorption lines have been shown to correlate with extinction due to interstellar dust \citep{richmond_ubvri_1994,munari_equivalent_1997}. In order to measure the extinction in the direction of SN~2021yja, we examine the \ion{Na}{1}\,D features in our Keck HIRES spectrum (pixel scale $\Delta x \approx 0.0026$~nm in the regions of interest), shown in Figure~\ref{fig:extinction}. The spectrum reveals four distinct absorption systems at $z=0.00002$ (Milky Way), 0.00559, 0.00566, and 0.00568.

We measure the equivalent widths of each of these eight lines by fitting and integrating a Gaussian profile, resulting in the values shown in Table~\ref{tab:ew}. (For the systems at $z=0.00566$ and 0.00568, we simultaneously fit two Gaussian profiles.) We then convert these equivalent widths to $E(B-V)$ using Eq.~9 of \cite{poznanski_empirical_2012} and applying the renormalization factor of 0.86 from \cite{schlafly_blue_2010}. The Milky Way extinction value, \EBVMW{}, is consistent with the value from \cite{schlafly_measuring_2011}, \EBVMWdust{}. We adopt the latter. The host galaxy extinction value (from the sum of the six lines) is \EBVhost{}.

\input{equivalent_widths}
\vspace{-12pt}

This is a surprisingly high value given the already very blue observed colors of SN~2021yja (see \S\ref{sec:colors}). To confirm this measurement, we also measure the equivalent width of the diffuse interstellar band (DIB) at 578~nm. From Eq.~6 of \cite{phillips_source_2013}, which can be rewritten
\begin{equation}
    W_\lambda = (0.059^{+0.039}_{-0.024}\mathrm{\ nm\ mag}^{-1}) E(B-V)
\end{equation}
using the extinction law of \cite{schlafly_measuring_2011}, we would expect an equivalent width of \EWDIB{}. Due to the lower signal-to-noise ratio of this feature, we fit a Gaussian with a fixed center (578~nm at $z=0.00568$) and width ($\mathrm{FWHM} = 0.222$~nm; \citealt{tuairisg_deep_2000}). Again integrating the best-fit Gaussian, we get an equivalent width consistent with our expectation, $W_\lambda = 0.0076$~nm. Given the more robust detection of the \ion{Na}{1}\,D lines, we adopt the extinction derived above, for a total of \EBVtot{}, for which we correct using the extinction law of \cite{fitzpatrick_correcting_1999}. Note that this only accounts for interstellar extinction in the host galaxy, and not for any extinction due to CSM.

\defcitealias{planckcollaboration_planck_2020}{Planck Collaboration (2020)}

\subsection{Distance}
Distance estimates for NGC~1325 based on the method of \cite{tully_new_1977} listed on the NASA/IPAC Extragalactic Database range from 17.7 to 26.1~Mpc ($\mu=31.24{-}32.09$~mag). For comparison, the redshift-dependent distance of NGC~1325 is $23.6 \pm 4.5$~Mpc, assuming the cosmological parameters of the \citetalias{planckcollaboration_planck_2020} and an uncertainty of $cz \approx 300\ \mathrm{km}\ \mathrm{s}^{-1}$ due to the galaxy's peculiar velocity.

As an additional check on the SN distance, we apply the expanding photosphere method \citep[EPM;][]{kirshner_distances_1974}, which is a geometrical technique that can independently constrain the distance of an individual Type~II SN. Assuming that the SN photosphere is expanding freely and spherically shortly after the explosion, we can derive the distance from the linear relation between the angular radius and the expanding velocity of the photosphere using the function
\begin{equation}
\label{equ:linear}
t = D\left(\frac{\theta}{v_{phot}}\right) + t_0
\end{equation}
where $D$ is the distance, $v_\mathrm{phot}$ and $\theta$ are the velocity and angular radius of the photosphere, respectively, and $t_0$ is the explosion epoch.
We estimate the photospheric velocity by measuring the minimum of the P~Cygni profile of the \ion{Fe}{2} line at 516.9~nm, which becomes detectable ${\sim}19$~days after explosion.
These measurements are listed in Table~\ref{tab:epm}.
In practice, the SN photosphere radiates as a dilute blackbody, so a dilution factor has to be involved when determining $\theta$ \citep{eastman_atmospheres_1996,hamuy_distance_2001,dessart_distance_2005}.
We combine the multiband photometry to simultaneously derive the angular size ($\theta$) and color temperature ($T_\mathrm{c}$) by minimizing the equation
\begin{equation}
\epsilon = \sum_{\nu\in S} \{ m_\nu + 5log[\theta\xi(T_c) ] - A_\nu - b_\nu(T_c)\}^2,
\end{equation}
where $\xi$ and $b_\nu$ are the dilution factor and the synthetic magnitude, respectively, both of which can be treated as a function of $T_\mathrm{c}$ \citep{hamuy_distance_2001,dessart_distance_2005}, $A_\nu$ is the reddening, $m_\nu$ is the observed magnitude, and $S$ is the filter subset, i.e., $\{B,V\}$, $\{B,V,I\}$ and $\{V,I\}$. We use the dilution factors calculated by \cite{jones_distance_2009} with the atmosphere model from \cite{dessart_distance_2005}.
The best-fitting parameters
$D$ and $t_{0}$ in Eq.~\ref{equ:linear} are estimated with a Markov Chain Monte Carlo (MCMC) method, and the priors of parameters are assumed to be uniform. The upper bound on the prior of $t_{0}$ is set to be MJD 59464.63, the first detection of the SN. \cite{jones_distance_2009} showed that there is a clear departure from linearity between $\theta/ v$ and $t$ ${\gtrsim}40$ days after explosion. Therefore, only data before this phase are used in our calculations (see Table~\ref{tab:epm}). For the three filter subsets, we obtain distances of $24.0^{+2.5}_{-1.4}$~Mpc, $22.4^{+0.8}_{-0.5}$~Mpc, and $23.5^{+2.2}_{-1.3}$~Mpc, respectively. The explosion epochs are constrained to be $-2.2^{+1.7}_{-3.2}$~d, $-0.6^{+0.6}_{-1.0}$~d, and $-2.1^{+1.6}_{-3.0}$~d, respectively, relative to the estimate from our shock cooling model (see \S\ref{sec:sc}).

Seeing that these distances are consistent with the Tully-Fisher distance above, and given the large uncertainties from this method, we adopt the most recent Tully-Fisher estimate (calculated using the $I$ band), $23.4^{+5.4}_{-4.4}$~Mpc \citep[$\mu = 31.85 \pm 0.45$~mag;][]{tully_cosmicflows-3_2016}. However, we keep in mind that all of our luminosity-dependent measurements suffer from a large systematic uncertainty in the distance.

\movetabledown=2.2in
\begin{rotatetable*}
\input{SN2021yja_EPM_results}
\end{rotatetable*}

\section{Analysis}\label{sec:analysis}
\subsection{Bolometric Light Curve and Colors\label{sec:colors}}
\begin{figure}
    \centering
    \includegraphics[width=\columnwidth]{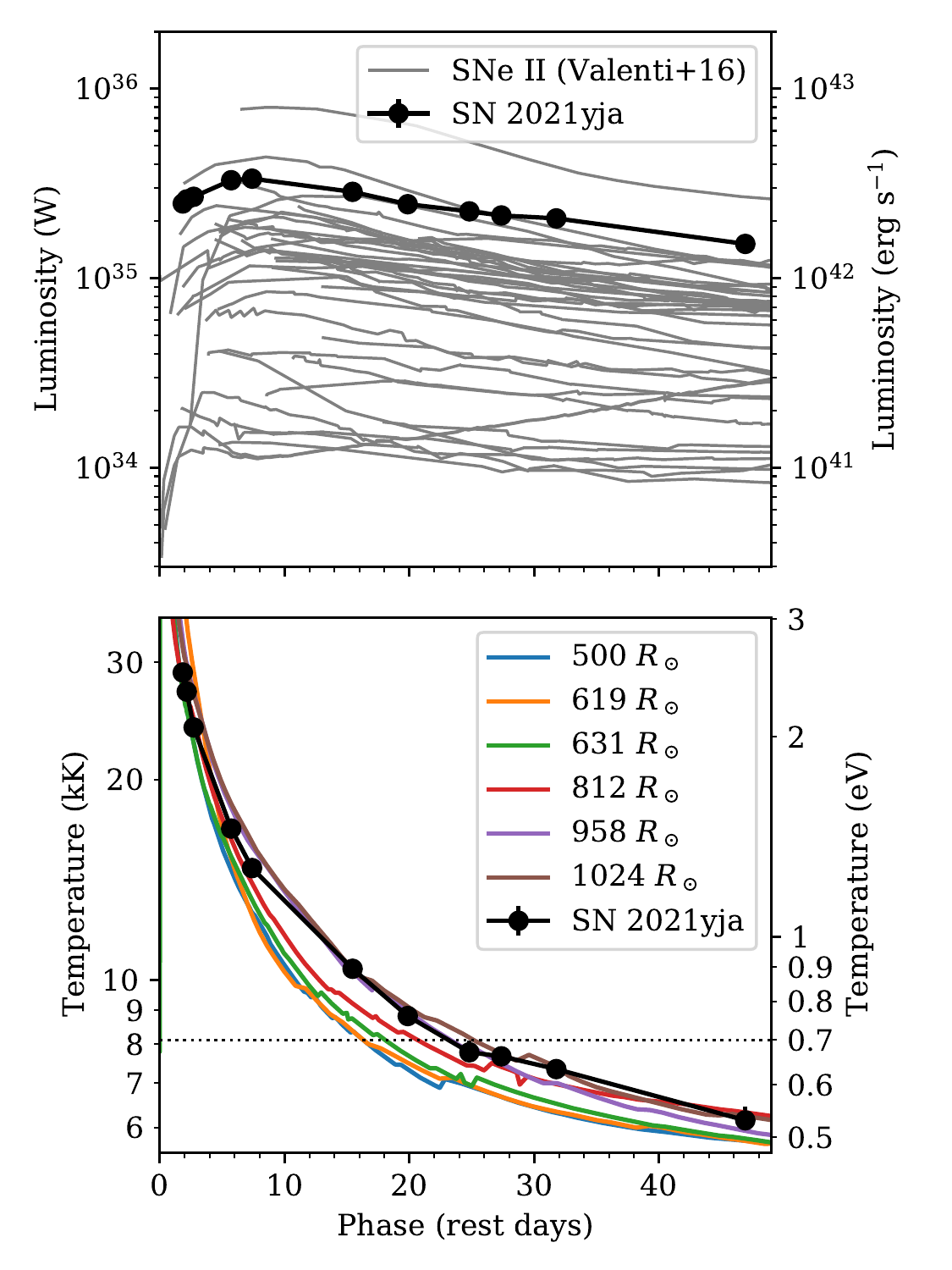}
    \caption{Top: Pseudobolometric light curve of SN~2021yja (black) compared to a sample of 39 pseudobolometric light curves from \cite{valenti_diversity_2016}. Bottom: Evolution of the blackbody temperature of SN~2021yja compared to temperature evolution models of \cite{kozyreva_shock_2020}. SN~2021yja lies between the models with progenitor radii of $800~R_\sun$ and $1000~R_\sun$, although no model is significantly hotter than SN~2021yja. Note that the blackbody temperature does not cool below the hydrogen recombination threshold (${\sim}0.7$~eV; dotted line) for several weeks, seemingly inconsistent with the appearance of strong hydrogen P~Cygni lines after the first few days. This becomes relevant for the validity of our shock cooling model (\S\ref{sec:sc}). (The data used to create this figure are available.)}
    \label{fig:bolometric}
\end{figure}

For the purposes of constructing a bolometric light curve, we restrict ourselves to epochs where we have Swift observations, as UV photometry is critical for constraining the SED of young SNe. We supplement the Swift photometry with ground-based \textit{griz} observations taken within 0.5 days of the Swift observation window. We fit a blackbody spectrum to the observed SED at each epoch using an MCMC routine implemented in the Light Curve Fitting package \citep{hosseinzadeh_light_2020}. This gives the temperature and radius evolution in the bottom panel of Figure~\ref{fig:bolometric}. We then integrate the best-fit blackbody from the blue edge of the $U$ band to the red edge of the $I$ band to produce a pseudobolometric light curve, such that it will be comparable to pseudobolometric light curves in the literature. The top panel of Figure~\ref{fig:bolometric} compares the pseudobolometric light curve of SN~2021yja to 39 other SNe~II from \cite{valenti_diversity_2016}. SN~2021yja is among the most luminous.

A comparison with the light-curve model grid from \cite{hiramatsu_luminous_2021} suggests that a hydrogen-rich envelope mass ${\gtrsim}5\ M_\sun$ is required to reproduce the unusually long plateau duration (${\approx}140$~days). A higher explosion energy (${\gtrsim}2 \times 10^{51}$~erg) is also likely required to reproduce the plateau luminosity of SN~2021yja, which is higher than the models, suggesting an even more massive hydrogen-rich envelope according to light-curve scaling relations \citep[e.g.,][]{popov_analytical_1993,kasen_type_2009,goldberg_inferring_2019}.

As suggested by \citet{rubin_exploring_2017} and \citet{kozyreva_shock_2020}, the temperature evolution prior to the recombination phase can serve as a solid diagnostic of the progenitor radius. The temperature declines as a power law during this period and changes to a shallower decline after the onset of recombination \citep{nakar_early_2010,shussman_type_2016,sapir_uv/optical_2017,faran_recombination_2019}, making this phase relatively easy to distinguish. In the bottom panel of Figure~\ref{fig:bolometric}, we compare the temperature evolution of SN~2021yja to temperature models from \cite{kozyreva_shock_2020} for various progenitor radii. We find that SN~2021yja lies between the $800~R_\sun$ and $1000~R_\sun$ models. Although there is no larger/hotter model, this comparison demonstrates that the progenitor of SN~2021yja is at least consistent with typical RSG radii \citep[$100{-}1500~R_\sun$;][]{levesque_astrophysics_2017}, in contrast to our findings in \S\ref{sec:sc}.

\begin{figure*}
    \centering
    \includegraphics[width=0.8\textwidth]{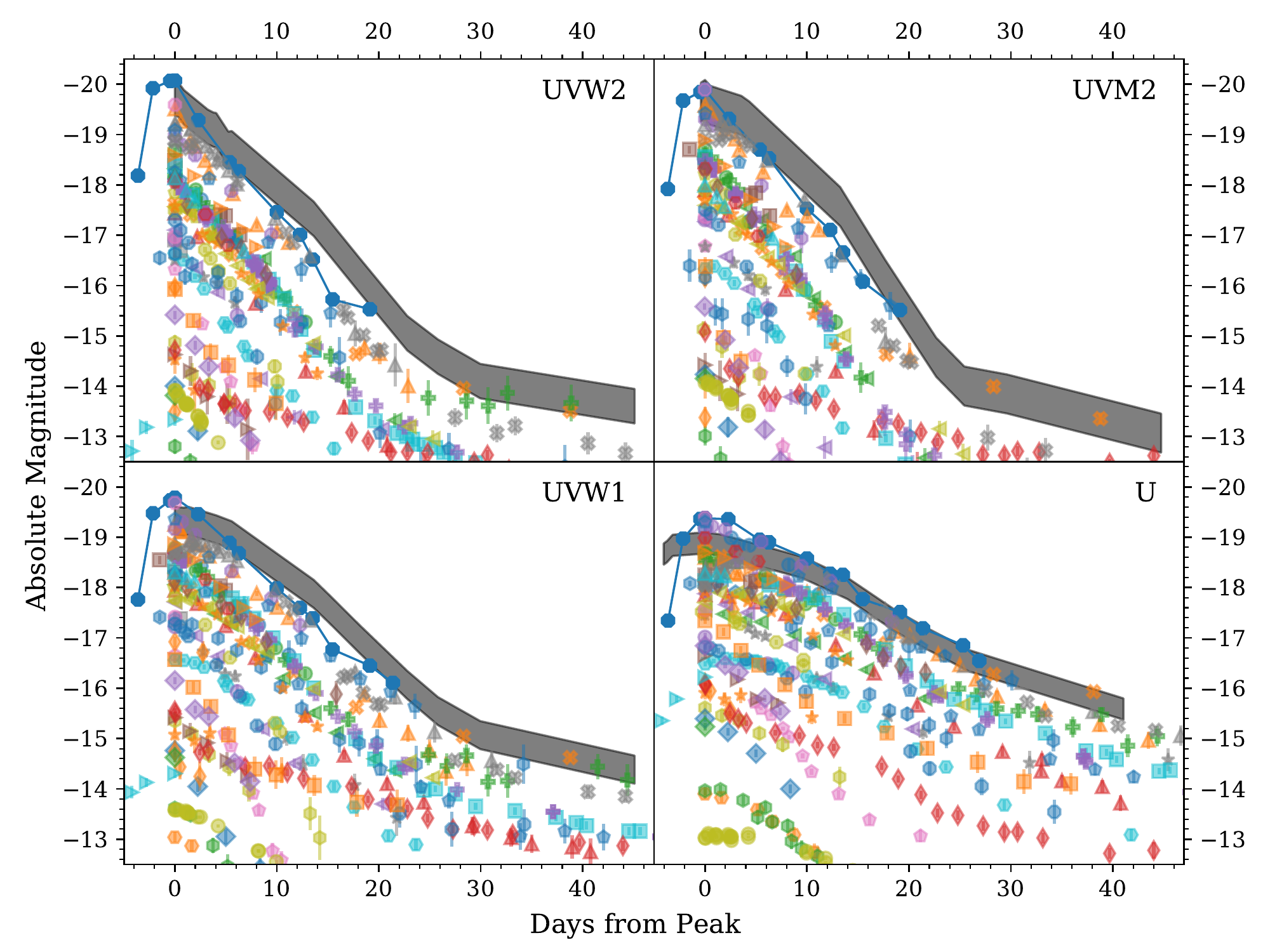}
    \caption{The UV light curve of SN~2021yja (shaded gray) compared to all SNe~II in the Swift Optical/Ultraviolet Supernova Archive (colored points; corrected for Milky Way extinction only; \citealt{brown_sousa:_2014}). All Swift data are in Vega magnitudes. The top of the gray region corresponds to \EBVhost{}, whereas the bottom is only corrected for Milky Way extinction. This makes SN~2021yja the second most UV-luminous SN~II observed by Swift, after SN~2020pni (blue octagons; \citealt{terreran_early_2022}). This unusually high UV luminosity suggests circumstellar interaction.}
    \label{fig:uv}
\end{figure*}

We also downloaded UV light curves of all SNe~II in the Swift Optical/Ultraviolet Supernova Archive \citep{brown_sousa:_2014}. We correct these for Milky Way extinction only and plot them in Figure~\ref{fig:uv}. SN~2021yja is the second most UV-luminous SN~II observed by Swift, after only SN~2020pni \citep{terreran_early_2022}.

\begin{figure*}
    \centering
    \includegraphics[width=0.7\textwidth]{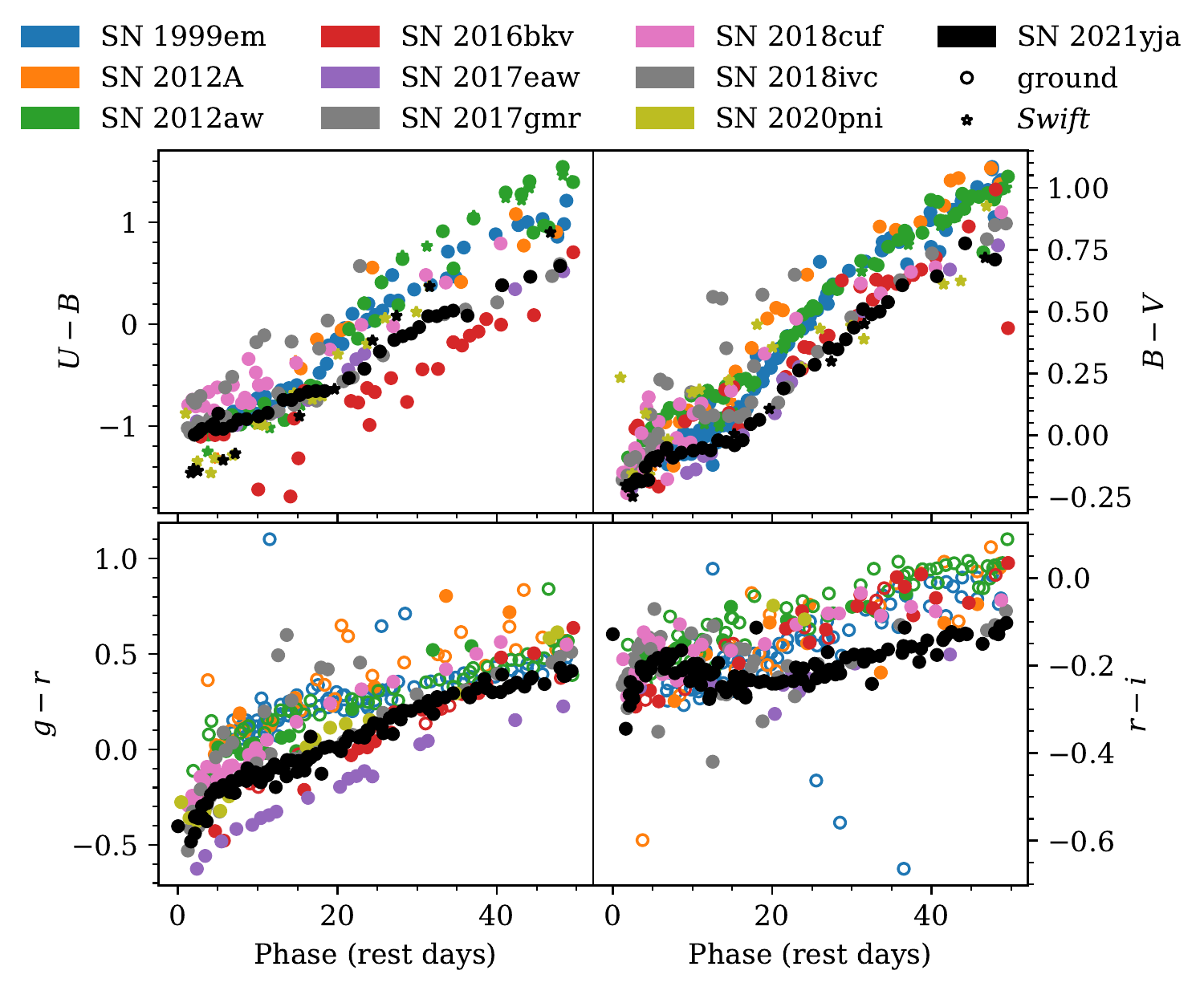}
    \caption{Extinction-corrected color curves of SN~2021yja compared to other SNe~II. Open markers for $g-r$ and $r-i$ are converted from $V-R$ and $R-I$, respectively, according to the relationships of \cite{jordi_empirical_2006}. The bump in the color curves ending around 15~d happens around the same time that a change in plateau slope is commonly observed in SNe~II \citep{anderson_characterizing_2014}, although such a change is not obvious in our light curve. Otherwise, the color evolution of SN~2021yja is fairly typical. (The data used to create this figure are available.)}
    \label{fig:colors}
\end{figure*}

Lastly, in Figure~\ref{fig:colors}, we compare the optical colors of SN~2021yja to a sample of other SNe~II \citep{hamuy_distance_2001,leonard_distance_2002,elmhamdi_photometry_2003,bose_supernova_2013,munari_bvri_2013,anderson_characterizing_2014,brown_sousa:_2014,dall'ora_type_2014,faran_photometric_2014,galbany_ubvriz_2016,rubin_type_2016,hosseinzadeh_short-lived_2018,dejaeger_berkeley_2019,szalai_type_2019,andrews_sn_2019,bostroem_discovery_2020,dong_supernova_2021,terreran_early_2022}. The evolution is fairly typical, starting near the minimum colors for a hot blackbody, then evolving redward, with an inflection point around 15 days after explosion. In the terminology of \cite{anderson_characterizing_2014}, this is the transition time $t_\mathrm{tran}$ when the light-curve slope changes from $s_1$ (the initial decline) to $s_2$ (the ``plateau''; see their Figure~1). Notably, we observe this transition in the colors even though the bolometric light curve in Figure~\ref{fig:bolometric} does not have a significant inflection point. We discuss this further in Section~\ref{sec:csm}.

\input{priors}
\vspace{-24pt}

\subsection{Shock Cooling Model}\label{sec:sc}
\begin{figure*}
    \centering
    \includegraphics[width=\textwidth]{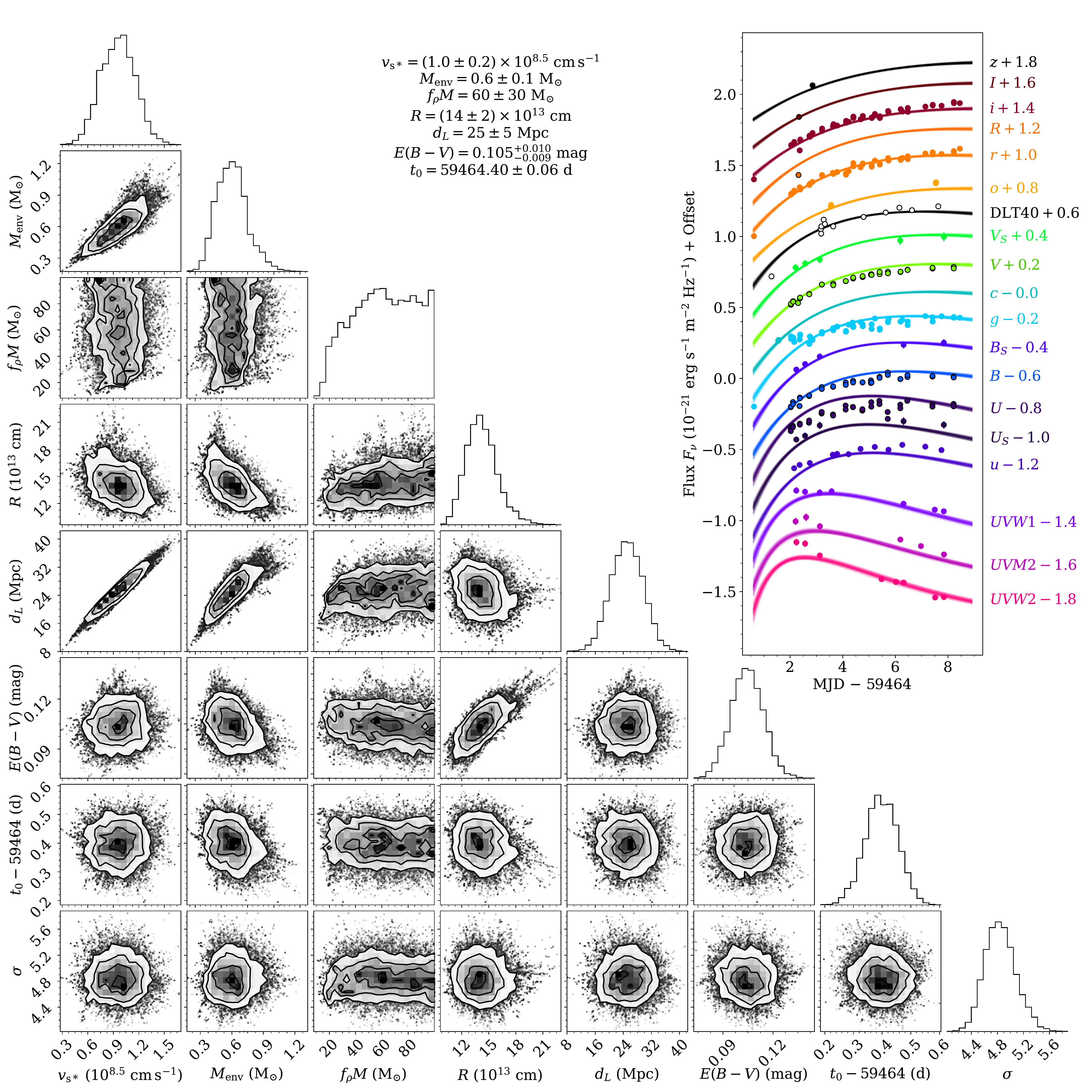}
    \caption{Modeling the shock cooling emission of SN~2021yja using the model of \cite{sapir_uv/optical_2017}. The parameters are described in Table~\ref{tab:prior}. We find that the first detection with MuSCAT3 occurred \deltaexplosion{} after the best-fit explosion time. We also find the best-fit progenitor radius to be \radius{}. This unusually large radius for an RSG can be lowered if we adopt a smaller extinction value or if we attribute some of the UV flux to interaction with CSM. (The data used to create this figure are available.)}
    \label{fig:shockcooling}
\end{figure*}

In the absence of circumstellar interaction, the early light curves of CCSNe are dominated by shock cooling emission, which depends upon the parameters of the explosion and the progenitor star, in particular the progenitor radius. We fit the model of \cite{sapir_uv/optical_2017} to the early light curve of SN~2021yja using the MCMC routine implemented in the Light Curve Fitting package \citep{hosseinzadeh_light_2020}. Because of the large uncertainties in extinction and distance outlined above, we allowed these to be free parameters in our model, with Gaussian priors according to their known values. We also included an intrinsic scatter term, $\sigma$, such that the effective uncertainty on each point is increased by a factor of $\sqrt{1+\sigma^2}$, with a half-Gaussian prior. This allows us to capture additional uncertainties, e.g., from photometric calibration, deviations from a blackbody spectrum, or small variations around the model, in our reported parameter uncertainties. The prior shapes and limits are given in Table~\ref{tab:prior}. We ran 100 walkers for 1000 steps to reach convergence, assessed via visual inspection of the chain history, and for an additional 1000 steps to sample the posterior. Figure~\ref{fig:shockcooling} shows the best-fit light curve, as well as the posterior distributions of and correlations between each parameter.

Qualitatively, the model fits SN~2021yja reasonably well across all bands. Unlike for several recent SNe~II that have a significant peak in their early light curve \citep[see, e.g.,][]{hosseinzadeh_short-lived_2018}, the shock cooling model is able to reproduce the smooth rise to plateau over ${\sim}10$ days in the reddest bands. The model is nominally valid until \tmax{} (\tmaxphase{} after explosion), at which point the temperature drops below 0.7 eV and recombination breaks the assumption of constant opacity \citep{sapir_uv/optical_2017}. This phase is roughly consistent with the recombination limit we estimated in Figure~\ref{fig:bolometric}. However, there are clear hydrogen features in the spectra starting around 10 days, implying that recombination is occurring much earlier than this. For this reason, we do not fit our light curve after maximum light in the reddest bands (i.e., we only fit the data shown in Figure~\ref{fig:shockcooling}). This inconsistency in recombination time is one clue that the physics behind the shock cooling model may not completely describe the data, despite the appearance of a good fit.

Putting this concern aside for the moment, the model is most sensitive to the progenitor radius. The best-fit progenitor radius is \radius{}, which is significantly larger than typical RSG radii \citep[$100{-}1500~R_\sun$;][]{levesque_astrophysics_2017}, and larger than the radius we infer in \S\ref{sec:colors}. Inspecting the correlations in Figure~\ref{fig:shockcooling}, we see that the radius estimate is strongly (positively) correlated with the extinction estimate. In other words, if we decrease our adopted extinction by $1{-}2\sigma$, we would estimate a more reasonable radius. However, we performed a second fit with a uniform prior in $E(B-V)$ and derived a best-fit extinction value similar to that in \S\ref{sec:extinction}. (As \citealt{sapir_uv/optical_2017} assume a blackbody SED in their model, this is equivalent to determining the extinction required to redden a blackbody into the observed SED.) Alternatively, as we discuss further in Section~\ref{sec:csm}, the extreme UV flux may be the effect of circumstellar interaction on our light curve, which is not included in the shock cooling model. If we could model this interaction component separately, a smaller radius might be required to generate the shock cooling component. We can treat our best-fit radius as a robust upper limit on the true progenitor radius, though admittedly it is not very constraining.

We also find a best-fit explosion (shock-breakout) time of \explosion{}, a mere \deltaexplosion{} before our serendipitous detection with MuSCAT3. One possibility, first suggested by \cite{kilpatrick_at2021yja:_2021}, is that this first detection is of the progenitor prior to shock breakout. Indeed the absolute magnitude at this phase ($M_r=-11.5$~mag, $M_i=-11.3$~mag) is similar to the ``precursor emission'' of SN~2020tlf observed by \cite{jacobson-galan_final_2022}. To investigate this possibility, we performed the light-curve fit again without this first detection. The best-fit explosion time is still before the MuSCAT3 detection, and the model has a slightly larger intrinsic scatter ($\sigma \approx 5.8$). In the absence of extraordinary proof to the contrary, we conclude that this detection was after explosion and adopt the explosion time above as $\mathrm{phase}=0$ throughout our figures.

\subsection{Nickel Mass\label{sec:ni}}
Near the end of our observations presented here, as of around 138 rest-frame days after explosion, the light curve of SN~2021yja has settled on the radioactive-decay tail, where the $^{56}\mathrm{Co} \to ^{56}\mathrm{Fe}$ decay provides most of the input power. We confirm that the bolometric decline rate over the last four epochs of photometry from Las Cumbres Observatory is 0.0096~mag~day$^{-1}$, very close to the $^{56}\mathrm{Co} \to ^{56}\mathrm{Fe}$ decay rate (0.0098~mag~day$^{-1}$; \citealt{colgate_early_1969}). During this phase, we can measure the mass of radioactive $^{56}$Ni produced in the explosion by scaling the luminosity to that of SN~1987A, which had an independent $^{56}$Ni measurement \citep[$M_\mathrm{Ni,87A}=0.075\ M_\sun$;][]{spiro_low_2014,valenti_diversity_2016}.

We calculate the bolometric luminosity for these four epochs by directly integrating the SED over the $U$ to $i$ filters. This is different than the blackbody-fitting method we used during the photospheric phase, because at these late times the SED is no longer described by a \cite{planck_vorlesungen_1906} function \citep[e.g.,][]{martinez_type_2022}. We calculate a bolometric light curve of SN~1987A with the same method using photometry from \cite{catchpole_spectroscopic_1988} via the Open Supernova Catalog \citep{guillochon_open_2017}, interpolate to the four epochs of SN~2021yja photometry, and take the average ratio. We find that SN~2021yja produced $M_\mathrm{Ni} = 0.141^{+0.074}_{-0.049}\ M_\sun$, where the uncertainty is dominated by our distance uncertainty. This places SN~2021yja at the 97th percentile for nickel mass among SNe~II \citep{anderson_meta-analysis_2019}, although we caution that this measurement is only ${\approx}1.6\sigma$ above the nickel mass of SN~1987A. Another possibility is that our measurement is inflated by extra luminosity from ongoing circumstellar interaction, although the decline slope we measure above suggests that $^{56}$Co decay is the dominant power source. An unusually large nickel mass may suggest a massive progenitor, according to the correlations of \cite{eldridge_supernova_2019}. We discuss this possibility further in \S\ref{sec:progenitor}.

\subsection{Spectral Features}\label{sec:spec}
The spectroscopic evolution of SN~2021yja ${\gtrsim}6$~days after explosion is typical for a SN~II: P~Cygni lines of hydrogen and helium superimposed on a cooling blackbody. Metal lines, including \ion{Ba}{2} and \ion{Fe}{2} begin to appear several weeks after explosion, blanketing the blue side of the spectrum. Its infrared spectra (Figure~\ref{fig:spec_nir}) place it in the ``weak'' class of \cite{davis_carnegie_2019}, with weak helium absorption at 1.083 $\mu$m in our ${>}100$~d spectra, a high-velocity component of the same line, and more pronounced \ion{Sr}{2} absorption. As such, SN~2021yja is consistent with the correlation between slow plateau decline rates and weak infrared helium absorption \citep{davis_carnegie_2019}. No carbon monoxide emission is detected at 100 days after explosion.

However, our earliest optical spectra, taken around 2 days after explosion, are somewhat unusual in the sense that they do not show the narrow high-ionization lines seen in many SNe~II this early. These lines indicate short-lived circumstellar interaction, which ends when the nearby circumstellar material (CSM) is swept up by the expanding ejecta \citep{gal-yam_wolf-rayet-like_2014,smith_ptf11iqb:_2015,khazov_flash_2016,yaron_confined_2017,bullivant_sn_2018,hosseinzadeh_short-lived_2018,soumagnac_sn_2020,bruch_large_2021,hiramatsu_electron-capture_2021,terreran_early_2022}. Figure~\ref{fig:spec_early} compares the early spectra of SN~2021yja to other SNe~II with well-constrained explosion dates and high-quality spectra taken within 6 days of explosion. Some show narrow \ion{He}{2}, \ion{C}{3}, \ion{N}{3}, and other lines: SN~2013fs \citep{yaron_confined_2017,bullivant_sn_2018}, SN~2017ahn \citep{tartaglia_early_2021}, SN~2018zd \citep{hiramatsu_electron-capture_2021}, and SN~2020pni \citep{terreran_early_2022}. SN~2017gmr \citep{andrews_sn_2019} and SN~2021yja do not show these lines, but instead show a unique ``ledge-shaped'' feature around 450--480~nm, which we discuss below.

\begin{figure}
    \centering
    \includegraphics[width=\columnwidth]{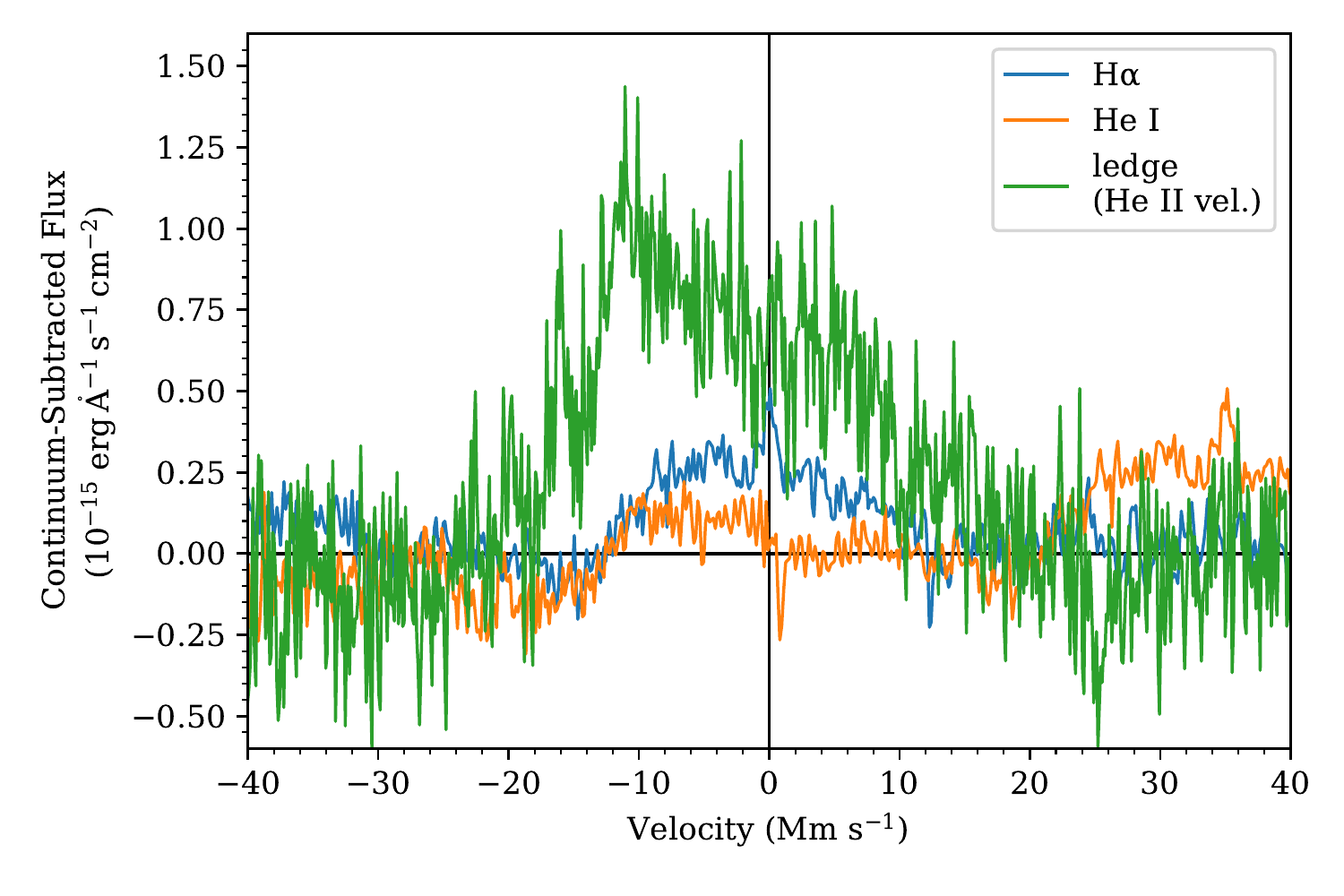}
    \caption{The profile of the unidentified feature around 450--480~nm (green) in the first spectrum of SN~2021yja, compared to the profiles of H$\alpha$ (blue) and He 587.6~nm (orange). \ion{He}{2} could be contributing at 468.6~nm, but the feature is broad and has an irregular shape, unlike the other P~Cygni profiles. As suggested by \cite{soumagnac_sn_2020} and \cite{bruch_large_2021}, we suspect this is a combination of several lines, possibly indicating a low level of circumstellar interaction. However, its breadth cannot be fully explained by the handful of narrow flash-ionized lines they suggest.}
    \label{fig:ledge}
\end{figure}

Figure~\ref{fig:ledge} shows a detailed view of the ledge-shaped feature in the earliest spectrum of SN~2021yja, compared to H$\alpha$ and He 587.6~nm in the same spectrum. There are two lines of thought about the identity of this feature in the literature, both related to circumstellar interaction. \citet[their Figure~20]{bullivant_sn_2018} and \citet[their Figure~18]{andrews_sn_2019} interpret it as very broad, blueshifted \ion{He}{2} 468.6~nm (although there is also a narrow component of this line present) produced in the outermost layers of the SN ejecta. However, the line profile is somewhat more symmetrical in these SNe than in SN~2021yja. \citet[their Figure~7]{soumagnac_sn_2020} and \citet[their Figure~5]{bruch_large_2021} interpret this feature as a blend of various flash-ionized lines from the CSM: \ion{N}{5} 460.4 and 462.0~nm, \ion{N}{2} 463.1 and 464.3~nm, \ion{C}{4} 465.8~nm, and \ion{He}{2} 468.6~nm.

Neither of these explanations is fully satisfactory. The former (single, broad line) is difficult to reconcile with the irregular profile of the feature, when other features have typical P~Cygni profiles. In addition, the implied velocity is nearly 30,000~km~s$^{-1}$ (as measured from the \ion{He}{2} rest wavelength to the blue edge of the feature), which is much higher than any other feature in the spectrum. The latter explanation (blend of several narrow emission lines) is inconsistent with the breadth and strength of the feature. We suspect that the feature is indeed a blend of several lines, possibly including \ion{He}{2}, \ion{C}{3}, and \ion{N}{3}, although these must be either much broader or much more numerous than the features suggested by previous work. We discuss one such possibility in \S\ref{sec:csm}.

\subsection{Limits on a Pre-explosion Counterpart}\label{sec:preexplosion}

\begin{figure*}
    \centering
    \includegraphics[width=\textwidth]{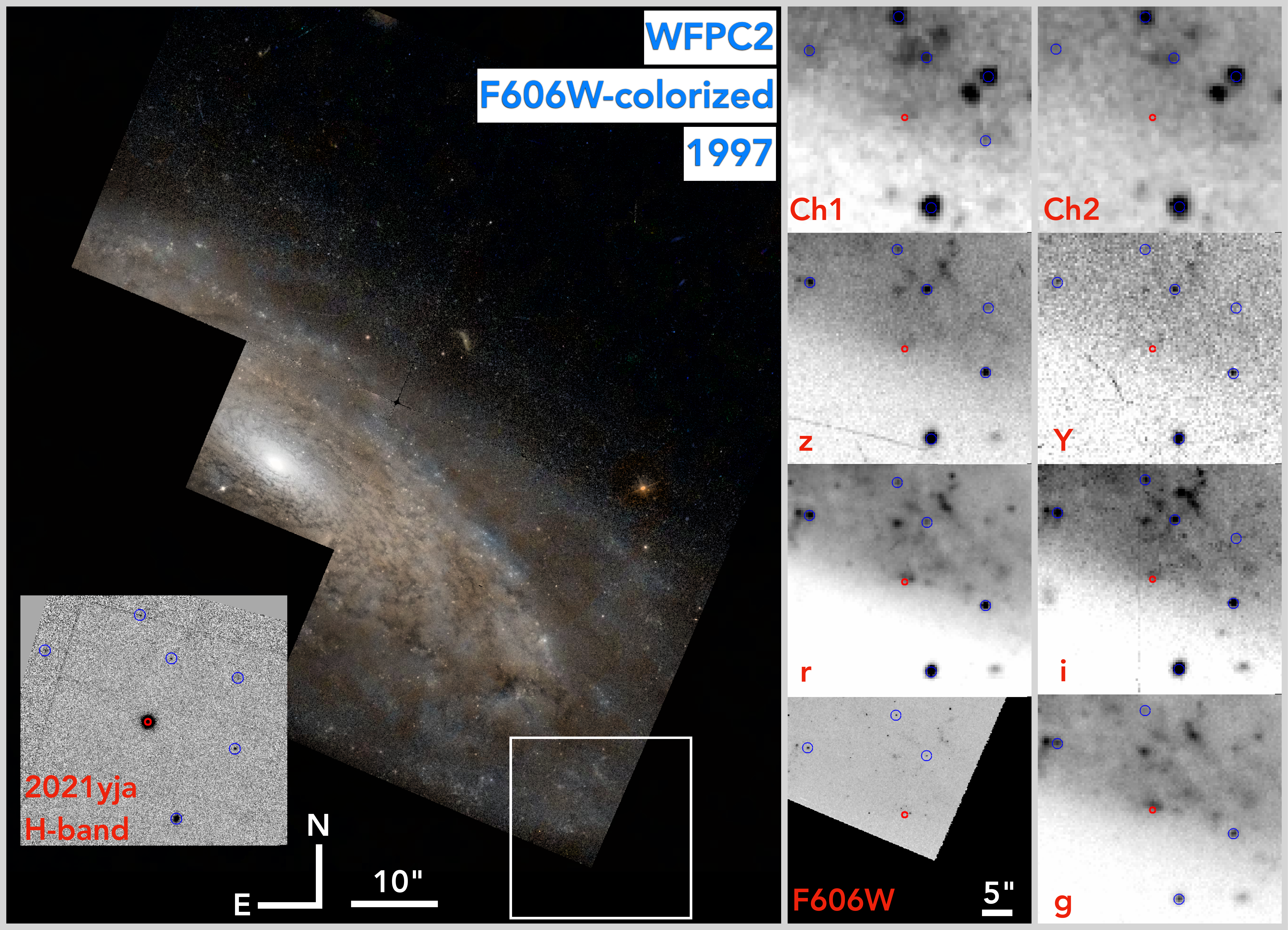}
    \caption{HST/WFPC2 F606W imaging of NGC\,1325 from 1997 colorized using DSS imaging of the same field (left) and showing the explosion site of SN\,2021yja (white square).  An inset panel shows GSAOI imaging of SN\,2021yja obtained in Dec.~2021 in a 40\arcsec$\times$40\arcsec\ region corresponding to the white frame.  The right-hand panels show the same region as seen in F606W, DECam $grizY$, and Spitzer/IRAC Channel 1 and 2, aligned using 3--6 common astrometric standards (blue circles) and centered on the site of SN~2021yja (red circle). The lack of a counterpart is described in \S\ref{sec:preexplosion}.}
    \label{fig:pre-imaging}
\end{figure*}

Figure~\ref{fig:pre-imaging} (left) shows our high-resolution GSAOI image of SN\,2021yja, aligned to the pre-explosion WFPC2 F606W image using three common astrometric sources in both frames, resulting in an rms astrometric uncertainty of 0.08\arcsec\ (0.8 WFPC2 pixels) between both frames.  In addition, we aligned the GSAOI image to DECam and Spitzer/IRAC imaging, using 4--6 common astrometric standards in both frames and resulting in an rms uncertainty of 0.07--0.12\arcsec\ (Figure~\ref{fig:pre-imaging}, right).

The position of SN\,2021yja does not align with any counterpart in the HST imaging.  The closest source is 0.88\arcsec\ (11$\sigma$) away from this position and detected at $m_{\rm F606W}=24.05\pm0.06$~mag \citep[previously discussed in][]{kilpatrick_at2021yja:_2021}.  The same is true in each of the other pre-explosion images as shown in Figure~\ref{fig:pre-imaging}, with no evidence for any significant, point-like emission within at least 3$\sigma$ of the site of SN\,2021yja.

Motivated by the depth of our imaging set, we place 3$\sigma$ limits on the presence of a counterpart in each of our pre-explosion images.  We estimate the standard deviation of the background variation within 2\arcsec\ of SN\,2021yja and use this value to estimate the maximum total flux that would not be flagged as a 3$\sigma$ detection within one PSF width at this location.  The resulting limits are given in AB mag in Table~\ref{tab:limits}, which we consider the 3$\sigma$ limits on the presence of any counterpart in each image.

\begin{deluxetable}{ccc}
\tablecaption{Limits on a Pre-explosion Counterpart to SN\,2021yja\label{tab:limits}}
\tablehead{\colhead{Instrument} & \colhead{Band} & \colhead{Magnitude}}
\startdata
WFPC2 & F606W & $>$27.15 \\
DECam & $g$   & $>$24.50 \\
DECam & $r$   & $>$24.08 \\
DECam & $i$   & $>$22.90 \\
DECam & $z$   & $>$22.56 \\
DECam & $Y$   & $>$21.90 \\
Spitzer/IRAC & Ch\,1 & $>$22.60 \\
Spitzer/IRAC & Ch\,2 & $>$22.20
\enddata
\tablecomments{All limits are given in AB magnitudes as described in \S\ref{sec:preexplosion}.}
\end{deluxetable}
\vspace{-12pt}

\begin{figure}
    \centering
    \includegraphics[width=\columnwidth]{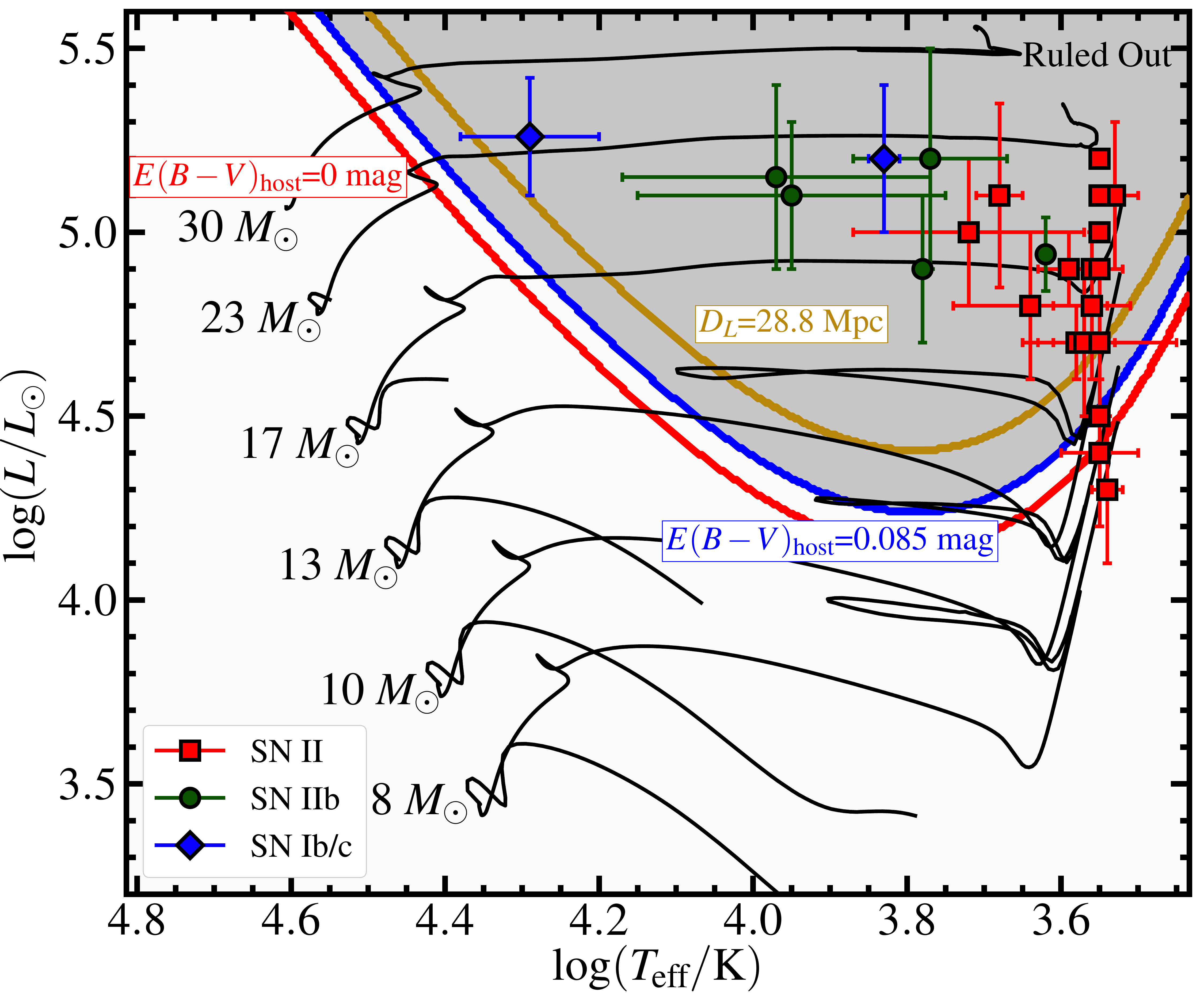}
    \caption{Hertzsprung-Russell diagram showing the region that is ruled out for a counterpart to SN\,2021yja by our limits in Table~\ref{tab:limits} (red line) and assuming $E(B-V)_{\rm host}=0.0$~mag and $D_{\rm host}=23.4$~Mpc.  For comparison, we show the same limits using the updated reddening value of $E(B-V)_{\rm host}=0.085$~mag and a total-to-selective extinction ratio of $R_{V}=3.1$ (blue), and the limits assuming this reddening and a farther distance of $28.8$~Mpc (gold; assumed distance + 1$\sigma$).  We also overplot MIST single-star evolutionary tracks for stars with initial masses 8--30~$M_{\odot}$ (black lines) as well as counterparts to Type Ib, IIb, and II SNe from the literature \citep[][and references therein]{cao_discovery_2013,smartt_observational_2015,kilpatrick_cool_2021}.  For our fiducial host distance and reddening values, our limits rule out evolved massive stars $>$9~$M_{\odot}$.}
    \label{fig:limits}
\end{figure}

To place these limits in context, we assume a blackbody spectral energy distribution with a given luminosity and show the region of the Hertzsprung-Russell diagram ruled out by our limits assuming a source at 23.4~Mpc and the extinction we derive in \S\ref{sec:extinction}.  We compare these limits to the expected luminosities and temperatures of single-star models derived from stars in MESA Isochrones \& Stellar Tracks \citep[MIST;][]{choi_mesa_2016}.  We can rule out all stars with initial masses $>$9~$M_{\odot}$ as shown in Figure~\ref{fig:limits}.  This analysis is dominated by the WFPC2 limit, which is significantly more constraining for the MIST models than DECam and Spitzer.  

Thus while we rely on the WFPC2 image from 26 March 1997, 24.5~yr before explosion, for our best constraints on the nature of any progenitor star of SN\,2021yja, we can place meaningful constraints on that star at later times when the DECam and Spitzer imaging were obtained.  These limits are informative in the scenario where that star was highly variable and had a lower flux in F606W compared to its average flux \citep[i.e., similar to variability in $\alpha$ Ori and the recent detection of a dimming RSG in M51 described in][]{levesque_betelgeuse_2020,jencson_exceptional_2022}.  We further explore the implications of this scenario below.

\defcitealias{h.e.s.s.collaboration_h.e.s.s._2022}{H.E.S.S. Collaboration (2022)}

\vspace{1in}
\section{Discussion}\label{sec:discuss}
\subsection{CSM}\label{sec:csm}
Our analysis above gives rise to a somewhat confusing picture with regard to CSM around the progenitor of SN~2021yja. On one hand, we do not see some standard signs of CSM, including pronounced, narrow emission lines in early spectra, and, at first glance, the shock cooling model appears to fit our observed light curve reasonably well. On the other hand, the high UV luminosity, very blue colors, and persistent high temperature (resulting in an overly large progenitor radius from the shock cooling model), as well as the ledge-shaped feature in the early spectra, do suggest some level of early circumstellar interaction. HST UV-optical spectra from \cite{vasylyev_early-time_2022} show no sign of circumstellar interaction, but these were observed much later (7 days after explosion) than our early spectra showing the ledge. In addition, the detection of delayed radio emission \citep{alsaberi_radio_2021} from SN~2021yja following an initial non-detection \citep{ryder_radio_2021} can be taken as an indication of interaction with a non-uniform CSM, similar to the conclusion of \cite{bostroem_signatures_2019} for the SN~II ASASSN-15oz. SN~2021yja was also observed in the very high-energy $\gamma$-ray domain ($E > 100$~GeV) by the High Energy Stereoscopic System (H.E.S.S.) Imaging Atmospheric Cherenkov Telescope Array, which will yield complementary constraints on the CSM density (H.E.S.S.\ Collaboration 2022, in preparation) following the work presented by \cite{abdalla_upper_2019} and the \citetalias{h.e.s.s.collaboration_h.e.s.s._2022}.

One way to resolve this apparent contradiction is to infer a small amount of CSM: less than is required to produce strong emission lines, but still enough to affect the broadband light curve, especially in the UV. We turn to the literature to constrain the mass-loss rate from both directions. By modeling a sample of early SN~II spectra with narrow emission lines, \cite{boian_progenitors_2020} find CSM density parameters of order $D \sim 10^{15{-}16}$~g~cm$^{-1}$, corresponding to mass loss of $\dot{M} \equiv 4 \pi v_\infty D \gtrsim 10^{-4}\ M_\sun\ \mathrm{yr}^{-1}$, where $v_\infty$ is the terminal wind speed. Our mass-loss rate must therefore be $\dot{M} \lesssim 10^{-4}\ M_\sun\ \mathrm{yr}^{-1}$.

\begin{figure*}
    \centering
    \includegraphics[width=\textwidth]{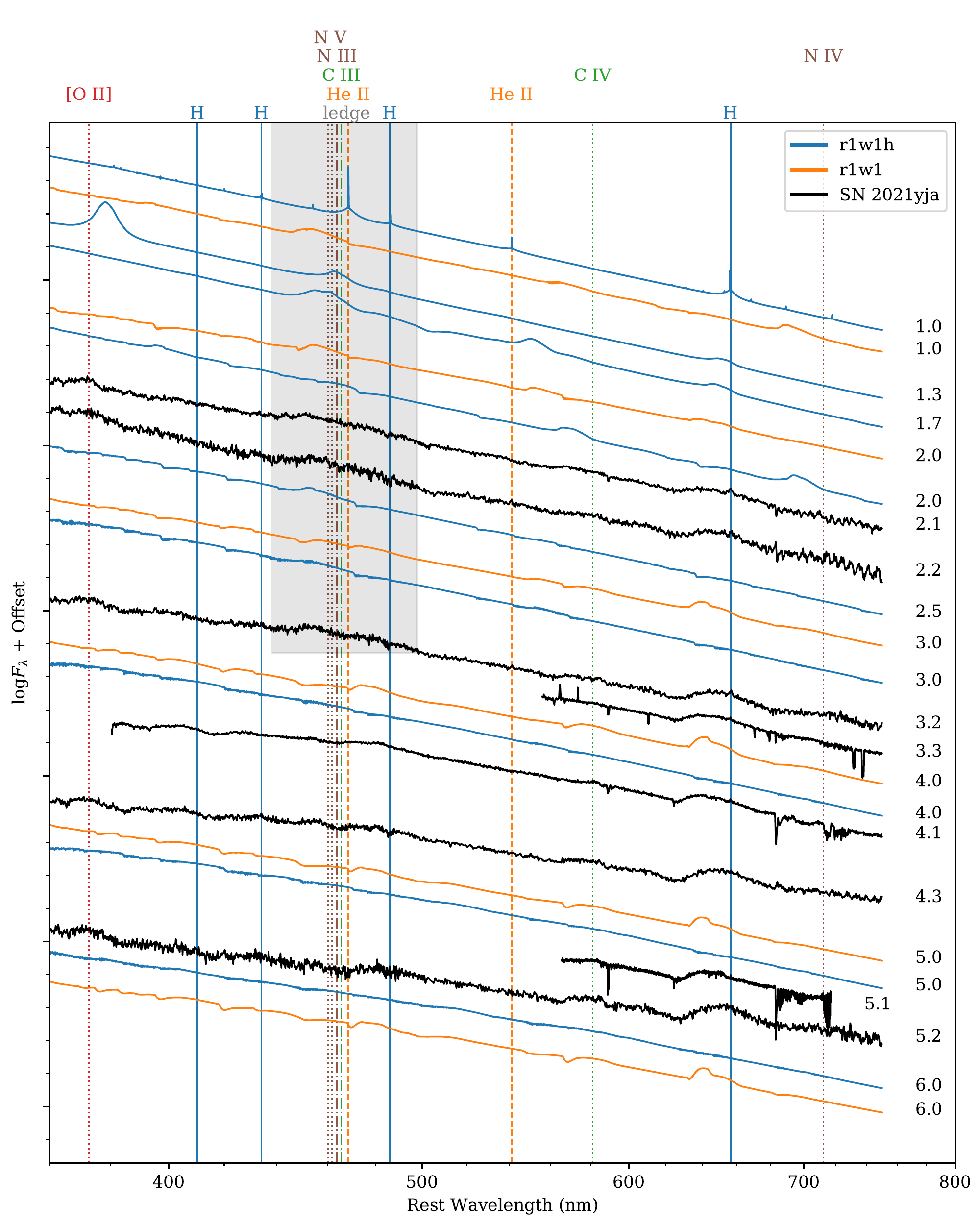}
    \caption{Comparing the early spectra of SN~2021yja to the \texttt{r1w1} and \texttt{r1w1h} models of \cite{dessart_explosion_2017}, with lines labeled as in Figure~\ref{fig:spec_early}. These models both have low-density CSM from a mass-loss rate of order $\dot{M} \approx 10^{-6}\ M_\sun\ \mathrm{yr}^{-1}$. The \texttt{r1w1h} model also includes an extended RSG atmosphere. Although not a perfect match to the models, the ledge-shaped feature we observe around 450--480~nm could be explained by weak circumstellar interaction, in particular by broad, blueshifted \ion{He}{2} and \ion{N}{5}.}
    \label{fig:models}
\end{figure*}

\cite{dessart_explosion_2017} propose a scenario in which the interaction lines are both broadened and blueshifted, which can potentially explain the ledge-shaped feature we discuss in \S\ref{sec:spec}. This occurs when an RSG with radius $R_\star = 501\ R_\sun$ explodes into low-density CSM ($\dot{M} = 10^{-6}\ M_\sun\ \mathrm{yr}^{-1}$). They term this the \texttt{r1w1} model. In this scenario, narrow emission lines are not seen or disappear within hours of explosion, at which time we had not yet observed the spectrum of SN~2021yja. When they add an extended atmosphere onto the RSG (\texttt{r1w1h} model; scale height $H_\rho = 0.3 R_\star$), the spectroscopic evolution is similar, but the light curve rises faster and is much more luminous in the UV (reaching $M_{UVW2} \approx -19.5$~mag). Such an extended atmosphere only requires a small energy injection during the last year before explosion \citep{smith_preparing_2014,morozova_influence_2020}. Figure~\ref{fig:models} compares the early spectra of SN~2021yja to the \texttt{r1w1} and \texttt{r1w1h} models of \cite{dessart_explosion_2017}. Notably, the extended atmosphere or \texttt{r1w1h} could also lead to an overestimate of the progenitor radius by a shock cooling model that was not designed to account for it. Therefore, we believe this to be the most likely scenario for the progenitor of SN~2021yja: an RSG with an extended atmosphere exploding into low-density CSM.

\subsection{Progenitor Star}\label{sec:progenitor}
Because of our skepticism about the inferred progenitor radius (\S\ref{sec:csm}), the best constraints we have on the progenitor are (1) the luminosity limit from direct imaging of the field before explosion and (2) the inferred nickel mass from the light-curve tail. Unfortunately, these are also somewhat in conflict with each other: the very strict luminosity limit suggests a very low-mass progenitor (with some caveats, discussed below), whereas the very large nickel mass suggests a massive progenitor \citep[${\gtrsim}20\ M_\sun$;][]{eldridge_supernova_2019}. The latter places SN~2021yja close to the upper limit for SN~II progenitors \citep{smartt_observational_2015,davies_initial_2018,davies_'on_2020,davies_'red_2020}, implying that its progenitor star ought to have $M_{V}=-5$ to $-6$~mag.

The limits from WFPC2 suggest that $M_{V}>-4.4$~mag, consistent with a progenitor star of initial mass ${<}9~M_{\odot}$ \citep[and following the single-star models of][]{choi_mesa_2016}.  We consider this a credible limit on the initial mass of the progenitor star derived using the same distance and extinction assumptions as SN\,2021yja above, but we acknowledge that there is some tension with the nature of the progenitor star as inferred from the SN light curve.  Our pre-explosion limit is dominated by the single epoch of WFPC2/F606W imaging from 1997, 24.5~years before explosion, which provides the deepest limits.  The tension between our pre-explosion constraints on the progenitor star and those derived from SN\,2021yja could be partly eased if the star were extinguished or variable on timescales of decades before core collapse.

A $-0.6$ to $-1.4$~mag discrepancy between this expectation and our derived limits could be explained by variability in the progenitor star $V$-band magnitude confined to a single epoch 24.5~yr prior to core collapse, similar to the variability observed in RSGs such as Betelgeuse and M51-DS1 \citep{levesque_betelgeuse_2020,jencson_exceptional_2022}.  The peak-to-peak variability in Betelgeuse was $\approx$1.2~mag such that even a fraction of this change in the SN\,2021yja progenitor star would completely obscure a $<$15~$M_{\odot}$ star in our WFPC2/F606W imaging.  We note, however, that these are extreme cases, and evidence from RSGs observed in M31 \citep{soraisam_variability_2018} and M51 \citep{conroy_complete_2018} suggests that the vast majority of RSGs in the expected luminosity range for the SN\,2021yja progenitor star do not exhibit such extreme variability.  Indeed, assuming that all RSGs follow similar $V$- and $I$-band variability to that observed in M51, those observed in \citet{conroy_complete_2018} suggest that $<$8\% of those stars will undergo peak-to-peak dimming at $M_{I}>-0.6$~mag.  However, this likelihood is highly uncertain for RSGs as a whole, and in particular, those that are $<$25~yr from core collapse, which may be even more variable than the overall population \citep[e.g.,][]{jacobson-galan_final_2022}. 

Alternatively, the CSM discussed above could significantly obscure the progenitor star in F606W if it contained a significant dust mass.  Assuming \cite{beasor_new_2020} mass-loss rates for an RSG in this mass range (${\approx}10^{-6}~M_\sun~\mathrm{yr}$) and corresponding wind speeds (20--25~km~s$^{-1}$), we consider the implied effect on circumstellar extinction.  Following methods in \citet{kilpatrick_connecting_2018} for a uniform wind with carbonaceous dust grains, we infer that SN\,2021yja would experience a line-of-sight extinction $A_{V}\approx 0.1$~mag, in which case circumstellar extinction could not explain an anomalously low optical luminosity for the SN\,2021ya progenitor star.  However, we note that this is strongly dependent on the assumed mass-loss rate and wind composition, which could be significantly different.

\section{Summary and Conclusions}
We have presented high-cadence, early photometric and spectroscopic observations of the Type~II SN~2021yja. These have conflicting interpretations with respect to the progenitor star and its CSM. On one hand, the light curve is very luminous, especially in the UV, but on the other hand, we see no narrow emission lines in the early spectra. The high luminosity on the radioactive-decay tail of the light curve also implies a large production of $^{56}$Ni (though with a large uncertainty), which suggests a large progenitor mass. However, we inspected archival HST imaging of the host galaxy of SN~2021yja and found no coincident sources that could be the progenitor. If stellar evolution models accurately describe the end state of stars, these limits suggest the progenitor was among the lowest-mass RSGs, ${\lesssim}9\ M_\sun$.

We conclude that the most likely progenitor is an RSG with an extended envelope, and that this progenitor exploded into low-density CSM produced via a mass-loss rate of order $\dot{M} \sim 10^{-6}\ M_\sun\ \mathrm{yr}^{-1}$. This CSM still has observable effects on the light curve and spectra. We emphasize that the details of the progenitor structure and CSM configuration must be considered in analyzing CCSNe; these lead to a multidimensional phase space of observables, i.e., a ``zoo'' of spectral features that do not map straightforwardly onto mass-loss rate or CSM density.

As we discover CCSNe increasingly early, thanks to specially designed survey strategies, detailed modeling of each well-observed event will allow us to piece together the statistics of the CCSN progenitor population. In particular, combining multiple independent lines of analysis, e.g., light-curve modeling, spectral modeling, and direct progenitor imaging, will allow us to build a complete picture of each new event, including mass-loss history, CSM density and composition, and progenitor structure. With a sample of well-studied events, we will gain a comprehensive view of the diversity of mass loss in massive stars in their final years.

\vspace{12pt}\noindent
We thank Luc Dessart for providing his model light curves and spectra and for insightful comments on the manuscript; Maria Jose Bustamante, Samaporn Tinyanont, C\'esar Rojas-Bravo, David Jones, and Kayla Loertscher for their effort in taking Kast data; UCSC undergraduates Cirilla Couch, Jessica Johnson, Payton Crawford, Srujan Dandu, and Zhisen Lai for their effort in taking Nickel data; Rosalie McGurk, Vivian U, and Tianmu Gao for obtaining the KCWI spectrum; and Andrew Howard and Howard Isaacson for reducing the HIRES spectrum.

Some of the data presented herein were obtained at the W.~M.~Keck Observatory, which is operated as a scientific partnership among the California Institute of Technology, the University of California, and the National Aeronautics and Space Administration. The Observatory was made possible by the generous financial support of the W.~M.~Keck Foundation. This research has made use of the Keck Observatory Archive (KOA), which is operated by the W.~M.~Keck Observatory and the NASA Exoplanet Science Institute (NExScI), under contract with the National Aeronautics and Space Administration. The authors wish to recognize and acknowledge the very significant cultural role and reverence that the summit of Maunakea has always had within the indigenous Hawai`ian community. We are most fortunate to have the opportunity to conduct observations from this mountain. 
Based on observations obtained at the international Gemini Observatory, a program of NSF's NOIRLab, which is managed by the Association of Universities for Research in Astronomy (AURA) under a cooperative agreement with the National Science Foundation. on behalf of the Gemini Observatory partnership: the National Science Foundation (United States), National Research Council (Canada), Agencia Nacional de Investigaci\'{o}n y Desarrollo (Chile), Ministerio de Ciencia, Tecnolog\'{i}a e Innovaci\'{o}n (Argentina), Minist\'{e}rio da Ci\^{e}ncia, Tecnologia, Inova\c{c}\~{o}es e Comunica\c{c}\~{o}es (Brazil), and Korea Astronomy and Space Science Institute (Republic of Korea).
Based on observations collected at the European Organisation for Astronomical Research in the Southern Hemisphere, Chile, as part of ePESSTO+ (the advanced Public ESO Spectroscopic Survey for Transient Objects Survey). ePESSTO+ observations were obtained under ESO programs ID 1103.D-0328 and 106.216C (PI: Inserra).
Based in part on observations obtained at the Southern Astrophysical Research (SOAR) telescope, which is a joint project of the Minist\'{e}rio da Ci\^{e}ncia, Tecnologia e Inova\c{c}\~{o}es (MCTI/LNA) do Brasil, the US National Science Foundation's NOIRLab, the University of North Carolina at Chapel Hill (UNC), and Michigan State University (MSU).
This publication has made use of data collected at Lulin Observatory, partly supported by MoST grant 109-2112-M-008-001.
B.E.T.\ acknowledges parts of this research were carried out on the traditional lands of the Ngunnawal people. We pay our respects to their elders past, present, and emerging. Based in part on data acquired at the Siding Spring Observatory 2.3\,m. We acknowledge the traditional owners of the land on which the SSO stands, the Gamilaraay people, and pay our respects to elders past and present.
Observations using Steward Observatory facilities were obtained as part of the large observing program AZTEC: Arizona Transient Exploration and Characterization.
We are grateful to the staff at Lick Observatory for their assistance with the Nickel telescope. Research at Lick Observatory is partially supported by a generous gift from Google.
The SALT data reported here were taken as part of Rutgers University program 2021-1-MLT-007 (PI: Jha).
This research is based on data obtained from the Astro Data Archive at NSF's NOIRLab. These data are associated with observing programs 2012B-0001 (PI J.~Frieman) and 2019B-1009 (PI P.~Lira).  NOIRLab is managed by the Association of Universities for Research in Astronomy (AURA) under a cooperative agreement with the National Science Foundation.
This research is based on observations made with the NASA/ESA Hubble Space Telescope obtained from the Space Telescope Science Institute, which is operated by the Association of Universities for Research in Astronomy, Inc., under NASA contract NAS 5-26555. These observations are associated with program GO-6359 \citep[PI Stiavelli;][]{hosseinzadeh_hst_2022}.
This work is based in part on observations made with the Spitzer Space Telescope, which was operated by the Jet Propulsion Laboratory, California Institute of Technology under a contract with NASA.
This work makes use of observations from the Las Cumbres Observatory network.

Time domain research by the University of Arizona team and D.J.S.\ is supported by NSF grants AST-1821987, 1813466, 1908972, \& 2108032, and by the Heising-Simons Foundation under grant \#2020-1864.
Research by Y.D., N.M., and S.V.\ is supported by NSF grants AST-1813176 and AST-2008108.
J.E.A.\ is supported by the international Gemini Observatory, a program of NSF's NOIRLab, which is managed by the Association of Universities for Research in Astronomy (AURA) under a cooperative agreement with the National Science Foundation, on behalf of the Gemini partnership of Argentina, Brazil, Canada, Chile, the Republic of Korea, and the United States of America.
K.A.B.\ acknowledges support from the DIRAC Institute in the Department of Astronomy at the University of Washington. The DIRAC Institute is supported through generous gifts from the Charles and Lisa Simonyi Fund for Arts and Sciences, and the Washington Research Foundation.
The Las Cumbres Observatory team is supported by NSF grants AST-1911225 and AST-1911151, and NASA Swift grant 80NSSC19K1639.
The UCSC team is supported in part by the Gordon and Betty Moore Foundation, the Heising-Simons Foundation, and by a fellowship from the David and Lucile Packard Foundation to R.J.F.
B.E.T.\ and his group  were supported by the Australian Research Council Centre of Excellence for All Sky Astrophysics in 3 Dimensions (ASTRO 3D), through project number CE170100013.
P.J.B.\ is partially supported by NASA Astrophysics Data Analysis grant NNX17AF43G ``Seeing Core-Collapse Supernovae in the Ultraviolet.''
C.A.\ and B.J.S.\ are supported by NASA grant 80NSSC19K1717 and NSF grants AST-1920392 and AST-1911074.
L.G.\ and T.E.M.B.\ acknowledge financial support from the Spanish Ministerio de Ciencia e Innovaci\'on (MCIN), the Agencia Estatal de Investigaci\'on (AEI) 10.13039/501100011033 under the PID2020-115253GA-I00 HOSTFLOWS project, from Centro Superior de Investigaciones Cient\'ificas (CSIC) under the PIE project 20215AT016, and by the program Unidad de Excelencia Mar\'ia de Maeztu CEX2020-001058-M. L.G.\ also acknowledges MCIN, AEI and the European Social Fund (ESF) ``Investing in your future'' under the 2019 Ram\'on y Cajal program RYC2019-027683-I.
M.G.\ is supported by the EU Horizon 2020 research and innovation program under grant agreement No.~101004719.
M.N.\ is supported by the European Research Council (ERC) under the European Union's Horizon 2020 research and innovation program (grant agreement No.~948381) and by a Fellowship from the Alan Turing Institute.

\facilities{ADS, ARC (DIS), ATT (WiFeS), Blanco (DECam), Bok (B\&C), CTIO:PROMPT, FLWO:1.2m (KeplerCam), FTN (FLOYDS, MuSCAT3), FTS (FLOYDS), Gemini:South (GSAOI), HST (WFPC2), Keck:I (HIRES), Keck:II (KCWI, NIRES), LCOGT (Sinistro), LO:1m (Lulin Compact Imager), Meckering:PROMPT, MMT (Binospec, MMIRS), NED, Nickel (Direct Imaging Camera), NTT (EFOSC, SOFI), OSC, SALT (RSS), Shane (Kast), SOAR (Goodman, TripleSpec), Spitzer (IRAC), Swift (UVOT), Thacher}

\software{Astropy \citep{astropycollaboration_astropy_2018}, BANZAI \citep{mccully_real-time_2018}, Binospec Pipeline \citep{kansky_binospec_2019}, corner \citep{foreman-mackey_corner.py:_2016}, DoPHOT \citep{schechter_dophot_1993}, emcee \citep{foreman-mackey_emcee_2013}, FLOYDS pipeline \citep{valenti_first_2014}, \texttt{hst123} \citep{kilpatrick_cool_2021}, IPython \citep{perez_ipython:_2007}, IRAF \citep{nationalopticalastronomyobservatories_iraf:_1999}, \texttt{lcogtsnpipe} \citep{valenti_diversity_2016}, Light Curve Fitting \citep{hosseinzadeh_light_2020}, KCWI Data Reduction Pipeline \citep{rizzi_keck_2021}, Matplotlib \citep{hunter_matplotlib:_2007}, MMIRS Pipeline \citep{chilingarian_data_2013}, NumPy \citep{oliphant_guide_2006}, \texttt{photpipe} \citep{rest_testing_2005}, POTPyRI (K.~Paterson et al.\ 2022, in preparation), PyKOSMOS \citep{davenport_pykosmos:_2021}, PyRAF \citep{sciencesoftwarebranchatstsci_pyraf:_2012}, PySALT 
\citep{crawford_pysalt:_2010}, PyWiFeS \citep{2014Ap&SS.349..617C}, QFitsView \citep{ott_qfitsview:_2012}, SciPy \citep{virtanen_scipy_2020}, \texttt{Spextool} \citep{cushing_spextool:_2004}, \texttt{xtellcor} \citep{vacca_method_2003}}

\appendix
\section{Photometry Reduction}\label{sec:photred}
Las Cumbres Observatory images are preprocessed by BANZAI \citep{mccully_real-time_2018}. We extracted aperture photometry from images taken under the Global Supernova Project using \texttt{lcogtsnpipe}, which is based on PyRAF \citep{sciencesoftwarebranchatstsci_pyraf:_2012}. We calibrated \textit{UBV} magnitudes to images of \cite{landolt_ubvri_1992} standard fields (Vega magnitudes) taken on the same night with the same telescope, and we calibrated \textit{gri} magnitudes to the Pan-STARRS1 (PS1) $3\pi$ Survey \citep[AB magnitudes;][]{chambers_pan-starrs1_2016}.

Las Cumbres images (including MuSCAT3) taken under the Young Supernova Experiment, as well as data from the Lulin 1\,m telescope and Nickel 1\,m telescope, were reduced using \texttt{photpipe} \citep{rest_testing_2005}. \textit{griz} magnitudes were calibrated to PS1 $3\pi$, and $u$ magnitudes were calibrated to SkyMapper \citep{onken_skymapper_2019}.

Thacher data were calibrated with nightly dark and bias frames and master flat-field frames in each band. We used DoPHOT \citep{schechter_dophot_1993} to perform PSF modeling and to obtain the counts in SN~2021yja as well as 12 reference stars in the Pan-STARRS1 catalog with $r$-band magnitudes between 13.7 and 16.6 mags. Preliminary magnitudes were calculated for SN~2021yja using zero-points derived for each image based on the reference star fluxes. Then a second pass was made on the data to perform a first order color correction to account for the bandpass mismatches between the Thacher and Pan-STARRS1 filters.

Unfiltered DLT40 light curves consist of aperture photometry calibrated to $r$-band catalogs from the AAVSO Photometry All-Sky Survey \citep[APASS;][]{henden_aavso_2009}.

Swift data were analyzed using an updated version of the pipeline for the SOUSA \citep{brown_sousa:_2014}. Zero-points from \cite{breeveld_updated_2011} were used with time-dependent sensitivity corrections updated in 2020 and an aperture correction calculated for 2021.

We perform image subtraction on each KeplerCam image with {\tt HOTPANTS} \citep{becker_hotpants:_2015}, using archival PS1 $3\pi$ images as reference templates. Instrumental magnitudes were measured using PSF fitting and calibrated to PS1 $3\pi$.

MMIRS imaging data were reduced using a custom pipeline (adapted from POTPyRI; K.~Paterson et al.\ 2022, in preparation) that performs standard dark-current subtraction, flat-fielding, sky background estimation and subtraction, astrometric alignments, and final stacking of the individual exposures. The images were photometrically calibrated with aperture photometry of relatively isolated stars in the MMIRS images with cataloged $JHK_\mathrm{s}$-band magnitudes in the Two Micron All Sky Survey (2MASS; \citealp{skrutskie_two_2006}). We then performed aperture photometry on SN\,2021yja. The measurement uncertainities are dominated by scatter in the zero-point estimation from the 2MASS stars. 

\section{Spectroscopy Reduction}\label{sec:specred}
The optical and near-infrared spectra of SN~2021yja are logged in Table~\ref{tab:spec}.
These will be made available on the Weizmann Interactive Supernova Data Repository \citep{yaron_wiserep_2012} after publication.

FLOYDS spectra were reduced using the FLOYDS pipeline \citep{valenti_first_2014}, which is based on PyRAF.

The Kast and Goodman spectra were reduced using standard IRAF/PyRAF and Python routines for bias/overscan subtractions and flat-fielding. The wavelength solution was derived using arc lamps while the final flux calibration and telluric lines removal were performed using spectrophotometric standard star spectra.

The EFOSC2 and SOFI spectra were reduced using the PESSTO pipeline \citep{smartt_pessto:_2015}.

Basic 2D reductions for the MMIRS standard long-slit $zJ$ single-order spectra were performed using the MMIRS data reduction pipeline \citep{chilingarian_data_2013}, including flat-fielding, subtraction of AB pairs to remove the sky background, and wavelength calibrations. We then performed 1D spectral extractions using standard tasks in IRAF. Flux calibrations and corrections for the strong near-infrared telluric absorption features were performed using observations of the A0V standards HD~13165, HIP~25347, and HIP~25117 taken immediately preceding each of the observations of the SN. We used the method of \citet{vacca_method_2003} implemented with the IDL tool \texttt{XTELLCOR\_GENERAL} developed by \citet{cushing_spextool:_2004} as part of the \texttt{Spextool} package. 

The SALT data were reduced using a custom pipeline based on the PySALT package \citep{crawford_pysalt:_2010}.

Each WiFeS observation was reduced using {\tt PyWiFeS} \citep{2014Ap&SS.349..617C} producing a 3D cube file for each grating that has had bad pixels and cosmic rays removed, while we combined our two 1800 exposures that we took on the night. We extracted spectra from the calibrated 3D data cube using QFitsView \citep{ott_qfitsview:_2012} using an aperture similar to the seeing on the night (average seeing of $1.1{-}1.6''$ on the night). For background subtraction, we extract a part of the sky that is isolated from the source using a similarly sized aperture.

Binospec data were reduced using the Binospec pipeline \citep{kansky_binospec_2019}.  While an internal flux calibration into relative flux units from throughput measurements of spectrophotometric standard stars is provided in the pipeline, we obtained our own spectrophotometric standard observations throughout the semester and created our own sensitivty function for flux calibration.

Reductions of Bok B\&C spectra were carried out using IRAF including bias subtraction, flat-fielding, and optimal extraction of the spectra. Flux calibration was achieved using spectrophotometric standards observed at an air mass similar to that of each science frame, and the resulting spectra were median combined into a single 1D spectrum for each epoch.

The DIS spectrum was reduced using PyKOSMOS \citep{davenport_pykosmos:_2021} using standard techniques for bias/overscan subtraction, flat-field correction, 1D extraction, and wavelength calibration. Flux calibration was performed using the optical flux standard star LTT2415. 1D observations were cleaned and median combined to produce the final spectrum.

The HIRES spectra were reduced using the standard procedure of the California Planet Search \citep{howard_california_2010}. We extracted the two orders of interest, containing the DIB and the \ion{Na}{1}\,D lines. For each order, we median combined the four exposures, masked the lines of interest, and then modeled the continuum by smoothing with a second-order \cite{savitzky_smoothing_1964} filter with a width of 555 pixels. Figure~\ref{fig:extinction} shows this spectrum divided by the continuum.

The KCWI spectra were reduced following the standard procedures used in the python version of the KCWI Data Reduction Pipeline \citep{rizzi_keck_2021}. The spectra were extracted from the data cube using QFitsView \citep{ott_qfitsview:_2012}.

The NIRES spectrum was reduced following standard procedures described in the IDL package \texttt{Spextool} version 5.0.2 for NIRES \citep{cushing_spextool:_2004}. The extracted 1D spectrum was flux calibrated and also corrected for telluric features with \texttt{xtellcorr} version 5.0.2 for NIRES \citep{vacca_method_2003} making use of observations of an A0V standard star.

We used the \texttt{Spextool} IDL package \citep{cushing_spextool:_2004} to reduce the TripleSpec data, and we subtracted consecutive AB pairs to remove the sky and the dark current. The pixel-to-pixel gain variations in the science frames were corrected by dividing for the normalized master flat. We calibrated 2D science frames in wavelength by using CuHeAr Hollow Cathode comparison lamps. To correct for telluric features and to flux-calibrate the SN spectra, we observed the HIP~19001 A0V telluric standard after the SN and at a similar air mass. After the extraction of the individual spectra from the 2D frames, we used the \texttt{xtellcorr} task \citep{vacca_method_2003} included in the \texttt{Spextool} IDL package to perform the telluric correction and the flux calibration.

\startlongtable
\input{spectroscopy}
\vspace{-36pt}

\bibliography{zotero_abbrev}
\bibliographystyle{aasjournal}

\end{document}

%% file: affil.tex
\newcommand{\LCO}{\affiliation{Las Cumbres Observatory, 6740 Cortona Drive, Suite 102, Goleta, CA 93117-5575, USA}}
\newcommand{\UCSB}{\affiliation{Department of Physics, University of California, Santa Barbara, CA 93106-9530, USA}}
\newcommand{\KITP}{\affiliation{Kavli Institute for Theoretical Physics, University of California, Santa Barbara, CA 93106-4030, USA}}
\newcommand{\UCD}{\affiliation{Department of Physics and Astronomy, University of California, Davis, 1 Shields Avenue, Davis, CA 95616-5270, USA}}
\newcommand{\WIS}{\affiliation{Department of Particle Physics and Astrophysics, Weizmann Institute of Science, 76100 Rehovot, Israel}}
\newcommand{\OKC}{\affiliation{Oskar Klein Centre, Department of Astronomy, Stockholm University, Albanova University Centre, SE-106 91 Stockholm, Sweden}}
\newcommand{\OAPD}{\affiliation{INAF -- Osservatorio Astronomico di Padova, Vicolo dell'Osservatorio 5, I-35122 Padova, Italy}}
\newcommand{\Caltech}{\affiliation{Cahill Center for Astronomy and Astrophysics, California Institute of Technology, Mail Code 249-17, Pasadena, CA 91125, USA}}
\newcommand{\GSFC}{\affiliation{Astrophysics Science Division, NASA Goddard Space Flight Center, Mail Code 661, Greenbelt, MD 20771, USA}}
\newcommand{\UMD}{\affiliation{Joint Space-Science Institute, University of Maryland, College Park, MD 20742, USA}}
\newcommand{\UCB}{\affiliation{Department of Astronomy, University of California, Berkeley, CA 94720-3411, USA}}
\newcommand{\TTU}{\affiliation{Department of Physics, Texas Tech University, Box 41051, Lubbock, TX 79409-1051, USA}}
\newcommand{\STScI}{\affiliation{Space Telescope Science Institute, 3700 San Martin Drive, Baltimore, MD 21218-2410, USA}}
\newcommand{\UT}{\affiliation{University of Texas at Austin, 1 University Station C1400, Austin, TX 78712-0259, USA}}
\newcommand{\IoA}{\affiliation{Institute of Astronomy, University of Cambridge, Madingley Road, Cambridge CB3 0HA, UK}}
\newcommand{\QUB}{\affiliation{Astrophysics Research Centre, School of Mathematics and Physics, Queen's University Belfast, Belfast BT7 1NN, UK}}
\newcommand{\IPAC}{\affiliation{Spitzer Science Center, California Institute of Technology, Pasadena, CA 91125, USA}}
\newcommand{\JPL}{\affiliation{Jet Propulsion Laboratory, California Institute of Technology, 4800 Oak Grove Dr, Pasadena, CA 91109, USA}}
\newcommand{\Southampton}{\affiliation{Department of Physics and Astronomy, University of Southampton, Southampton SO17 1BJ, UK}}
\newcommand{\LANL}{\affiliation{Space and Remote Sensing, MS B244, Los Alamos National Laboratory, Los Alamos, NM 87545, USA}}
\newcommand{\Tsinghua}{\affiliation{Physics Department and Tsinghua Center for Astrophysics, Tsinghua University, Beijing, 100084, People's Republic of China}}
\newcommand{\NAOC}{\affiliation{National Astronomical Observatory of China, Chinese Academy of Sciences, Beijing, 100012, People's Republic of China}}
\newcommand{\Itagaki}{\affiliation{Itagaki Astronomical Observatory, Yamagata 990-2492, Japan}}
\newcommand{\Einstein}{\altaffiliation{Einstein Fellow}}
\newcommand{\Hubble}{\altaffiliation{Hubble Fellow}}
\newcommand{\CfA}{\affiliation{Center for Astrophysics \textbar{} Harvard \& Smithsonian, 60 Garden Street, Cambridge, MA 02138-1516, USA}}
\newcommand{\UA}{\affiliation{Steward Observatory, University of Arizona, 933 North Cherry Avenue, Tucson, AZ 85721-0065, USA}}
\newcommand{\MPIA}{\affiliation{Max-Planck-Institut f\"ur Astrophysik, Karl-Schwarzschild-Stra\ss{}e 1, D-85748 Garching, Germany}}
\newcommand{\DSFP}{\altaffiliation{LSSTC Data Science Fellow}}
\newcommand{\HCO}{\affiliation{Harvard College Observatory, 60 Garden Street, Cambridge, MA 02138-1516, USA}}
\newcommand{\Carnegie}{\affiliation{Observatories of the Carnegie Institute for Science, 813 Santa Barbara Street, Pasadena, CA 91101-1232, USA}}
\newcommand{\TAU}{\affiliation{School of Physics and Astronomy, Tel Aviv University, Tel Aviv 69978, Israel}}
\newcommand{\Edinburgh}{\affiliation{Institute for Astronomy, University of Edinburgh, Royal Observatory, Blackford Hill EH9 3HJ, UK}}
\newcommand{\Birmingham}{\affiliation{Birmingham Institute for Gravitational Wave Astronomy and School of Physics and Astronomy,\\University of Birmingham, Birmingham B15 2TT, UK}}
\newcommand{\Bath}{\affiliation{Department of Physics, University of Bath, Claverton Down, Bath BA2 7AY, UK}}
\newcommand{\CTIO}{\affiliation{Cerro Tololo Inter-American Observatory, National Optical Astronomy Observatory, Casilla 603, La Serena, Chile}}
\newcommand{\Potsdam}{\affiliation{Institut f\"ur Physik und Astronomie, Universit\"at Potsdam, Haus 28, Karl-Liebknecht-Str. 24/25, D-14476 Potsdam-Golm, Germany}}
\newcommand{\INPE}{\affiliation{Instituto Nacional de Pesquisas Espaciais, Avenida dos Astronautas 1758, 12227-010, S\~ao Jos\'e dos Campos -- SP, Brazil}}
\newcommand{\UNC}{\affiliation{Department of Physics and Astronomy, University of North Carolina, 120 East Cameron Avenue, Chapel Hill, NC 27599, USA}}
\newcommand{\Ohio}{\affiliation{Astrophysical Institute, Department of Physics and Astronomy, 251B Clippinger Lab, Ohio University, Athens, OH 45701-2942, USA}}
\newcommand{\AAS}{\affiliation{American Astronomical Society, 1667 K~Street NW, Suite 800, Washington, DC 20006-1681, USA}}
\newcommand{\MMT}{\affiliation{MMT and Steward Observatories, University of Arizona, 933 North Cherry Avenue, Tucson, AZ 85721-0065, USA}}
\newcommand{\Geneva}{\affiliation{ISDC, Department of Astronomy, University of Geneva, Chemin d'\'Ecogia, 16 CH-1290 Versoix, Switzerland}}
\newcommand{\IUCAA}{\affiliation{Inter-University Center for Astronomy and Astrophysics, Post Bag 4, Ganeshkhind, Pune, Maharashtra 411007, India}}
\newcommand{\CMU}{\affiliation{Department of Physics, Carnegie Mellon University, 5000 Forbes Avenue, Pittsburgh, PA 15213-3815, USA}}
\newcommand{\NAOJ}{\affiliation{Division of Science, National Astronomical Observatory of Japan, 2-21-1 Osawa, Mitaka, Tokyo 181-8588, Japan}}
\newcommand{\IfA}{\affiliation{Institute for Astronomy, University of Hawai`i, 2680 Woodlawn Drive, Honolulu, HI 96822-1839, USA}}
\newcommand{\UCSC}{\affiliation{Department of Astronomy and Astrophysics, University of California, Santa Cruz, CA 95064-1077, USA}}
\newcommand{\Purdue}{\affiliation{Department of Physics and Astronomy, Purdue University, 525 Northwestern Avenue, West Lafayette, IN 47907-2036, USA}}
\newcommand{\Princeton}{\affiliation{Department of Astrophysical Sciences, Princeton University, 4 Ivy Lane, Princeton, NJ 08540-7219, USA}}
\newcommand{\Moore}{\affiliation{Gordon and Betty Moore Foundation, 1661 Page Mill Road, Palo Alto, CA 94304-1209, USA}}
\newcommand{\Durham}{\affiliation{Department of Physics, Durham University, South Road, Durham, DH1 3LE, UK}}
\newcommand{\JHU}{\affiliation{Department of Physics and Astronomy, The Johns Hopkins University, 3400 North Charles Street, Baltimore, MD 21218, USA}}
\newcommand{\Toronto}{\affiliation{David A.\ Dunlap Department of Astronomy and Astrophysics, University of Toronto,\\ 50 St.\ George Street, Toronto, Ontario, M5S 3H4 Canada}}
\newcommand{\Duke}{\affiliation{Department of Physics, Duke University, Campus Box 90305, Durham, NC 27708, USA}}
\newcommand{\NCU}{\affiliation{Graduate Institute of Astronomy, National Central University, 300 Jhongda Road, 32001 Jhongli, Taiwan}}
\newcommand{\Columbia}{\affiliation{Department of Physics and Columbia Astrophysics Laboratory, Columbia University, Pupin Hall, New York, NY 10027, USA}}
\newcommand{\Flatiron}{\affiliation{Center for Computational Astrophysics, Flatiron Institute, 162 5th Avenue, New York, NY 10010-5902, USA}}
\newcommand{\CIERA}{\affiliation{Center for Interdisciplinary Exploration and Research in Astrophysics and Department of Physics and Astronomy, \\Northwestern University, 1800 Sherman Avenue, 8th Floor, Evanston, IL 60201, USA}}
\newcommand{\GeminiNorth}{\affiliation{Gemini Observatory, 670 North A`ohoku Place, Hilo, HI 96720-2700, USA}}
\newcommand{\GeminiSouth}{\affiliation{Gemini Observatory, NSF's National Optical-Infrared Astronomy Research Laboratory, Casilla 603, La Serena, Chile}}
\newcommand{\Keck}{\affiliation{W.~M.~Keck Observatory, 65-1120 M\=amalahoa Highway, Kamuela, HI 96743-8431, USA}}
\newcommand{\UW}{\affiliation{Department of Astronomy, University of Washington, 3910 15th Avenue NE, Seattle, WA 98195-0002, USA}}
\newcommand{\DiRAC}{\altaffiliation{DiRAC Fellow}}
\newcommand{\USask}{\affiliation{Department of Physics \& Engineering Physics, University of Saskatchewan, 116 Science Place, Saskatoon, SK S7N 5E2, Canada}}
\newcommand{\Thacher}{\affiliation{Thacher School, 5025 Thacher Road, Ojai, CA 93023-8304, USA}}
\newcommand{\Rutgers}{\affiliation{Department of Physics and Astronomy, Rutgers, the State University of New Jersey,\\136 Frelinghuysen Road, Piscataway, NJ 08854-8019, USA}}
\newcommand{\FSU}{\affiliation{Department of Physics, Florida State University, 77 Chieftan Way, Tallahassee, FL 32306-4350, USA}}
\newcommand{\Melbourne}{\affiliation{School of Physics, The University of Melbourne, Parkville, VIC 3010, Australia}}
\newcommand{\ASTROthreeD}{\affiliation{ARC Centre of Excellence for All Sky Astrophysics in 3 Dimensions (ASTRO 3D), Australia}}
\newcommand{\Stromlo}{\affiliation{Mt.\ Stromlo Observatory, The Research School of Astronomy and Astrophysics, Australian National University, ACT 2601, Australia}}
\newcommand{\NCPAS}{\affiliation{National Centre for the Public Awareness of Science, Australian National University, Canberra, ACT 2611, Australia}}
\newcommand{\TAMU}{\affiliation{Department of Physics and Astronomy, Texas A\&M University, 4242 TAMU, College Station, TX 77843, USA}}
\newcommand{\Mitchell}{\affiliation{George P.\ and Cynthia Woods Mitchell Institute for Fundamental Physics \& Astronomy, College Station, TX 77843, USA}}
\newcommand{\ESO}{\affiliation{European Southern Observatory, Alonso de C\'ordova 3107, Casilla 19, Santiago, Chile}}
\newcommand{\ICE}{\affiliation{Institute of Space Sciences (ICE, CSIC), Campus UAB, Carrer
de Can Magrans, s/n, E-08193 Barcelona, Spain}}
\newcommand{\IEEC}{\affiliation{Institut d'Estudis Espacials de Catalunya, Gran Capit\`a, 2-4, Edifici Nexus, Desp.\ 201, E-08034 Barcelona, Spain}}
\newcommand{\Warwick}{\affiliation{Department of Physics, University of Warwick, Gibbet Hill Road, Coventry CV4 7AL, UK}}
\newcommand{\Macquarie}{\affiliation{School of Mathematical and Physical Sciences, Macquarie University, NSW 2109, Australia}}
\newcommand{\AAARC}{\affiliation{Astronomy, Astrophysics and Astrophotonics Research Centre, Macquarie University, Sydney, NSW 2109, Australia}}
\newcommand{\Capodimonte}{\affiliation{INAF -- Capodimonte Astronomical Observatory, Salita Moiariello 16, I-80131 Napoli, Italy}}
\newcommand{\INFNNapoli}{\affiliation{INFN -- Napoli, Strada Comunale Cinthia, I-80126 Napoli, Italy}}
\newcommand{\ICRANet}{\affiliation{ICRANet, Piazza della Repubblica 10, I-65122 Pescara, Italy}}
\newcommand{\Warsaw}{\affiliation{Astronomical Observatory, University of Warsaw, Al.\ Ujazdowskie 4, 00-478 Warszawa, Poland}}
\newcommand{\IAASARS}{\affiliation{IAASARS, National Observatory of Athens, 15236 Penteli, Greece}}
\newcommand{\NKUA}{\affiliation{Department of Astrophysics, Astronomy \& Mechanics, Faculty of Physics,\\National and Kapodistrian University of Athens, 15784 Athens, Greece}}
\newcommand{\IAIFI}{\affiliation{The NSF AI Institute for Artificial Intelligence and Fundamental Interactions, USA}}
\newcommand{\Cardiff}{\affiliation{Cardiff Hub for Astrophysics Research and Technology, School of Physics \& Astronomy, Cardiff University, Queens Buildings, The Parade, Cardiff, CF24 3AA, UK}}

%% file: equivalent_widths.tex
\begin{deluxetable}{ccc}
\tablecaption{Equivalent Widths\label{tab:ew}}
\tablehead{\colhead{Line} & \colhead{Redshift} & \colhead{$W_\lambda$ (nm)}}
\startdata
DIB & 0.00568 & 0.0076 \\
\ion{Na}{1D$_2$} & 0.00002 & 0.0059 \\
\ion{Na}{1D$_1$} & 0.00002 & 0.0024 \\
\ion{Na}{1D$_2$} & 0.00559 & 0.0134 \\
\ion{Na}{1D$_2$} & 0.00566 & 0.0148 \\
\ion{Na}{1D$_2$} & 0.00568 & 0.0108 \\
\ion{Na}{1D$_1$} & 0.00559 & 0.0111 \\
\ion{Na}{1D$_1$} & 0.00566 & 0.0128 \\
\ion{Na}{1D$_1$} & 0.00568 & 0.0091
\enddata
\end{deluxetable}

%% file: SN2021yja_EPM_results.tex
\begin{deluxetable*}{cCCCC|CCCCCC}
\tablecaption{Data and Results for the Expanding Photosphere Method\label{tab:epm}}
\tablecolumns{11}
\tablehead{\colhead{Phase} & \colhead{\ion{Fe}{2} Velocity} & \colhead{$B$} & \colhead{$V$} & \colhead{$I$} & \twocolhead{$BV$} & \twocolhead{$BVI$} & \twocolhead{$VI$} \\[-12pt]
\colhead{} & \colhead{} & \colhead{} & \colhead{} & \colhead{} & \twocolhead{------------------------------------} & \twocolhead{------------------------------------} & \twocolhead{------------------------------------} \\[-12pt]
\colhead{(d)} & \colhead{(Mm s$^{-1}$)} & \colhead{(mag)} & \colhead{(mag)} & \colhead{(mag)} & \colhead{$\theta / v$ (Mpc$^{-1}$ d)} & \colhead{Temp. (K)} & \colhead{$\theta / v$ (Mpc$^{-1}$ d)} & \colhead{Temp. (K)} & \colhead{$\theta / v$ (Mpc$^{-1}$ d)} & \colhead{Temp. (K)}}
\startdata
19.0 & 8.122 \pm 0.389 & 14.733 \pm 0.056 & 14.549 \pm 0.062 & 14.256 \pm 0.045 & 0.805 \pm 0.127 & 11533^{+2103}_{-1500} & 0.847 \pm 0.068 & 10735^{+640}_{-573} & 0.866 \pm 0.123 & 10438^{+1308}_{-1004} \\
22.0 & 8.091 \pm 0.161 & 14.894 \pm 0.061 & 14.582 \pm 0.054 & 14.201 \pm 0.029 & 0.940 \pm 0.100 & 9353^{+1260}_{-989} & 0.990 \pm 0.044 & 9213^{+395}_{-357} & 1.015 \pm 0.087 & 9258^{+759}_{-632} \\
23.0 & 7.747 \pm 0.180 & 14.894 \pm 0.061 & 14.548 \pm 0.063 & 14.192 \pm 0.028 & 1.032 \pm 0.113 & 8907^{+1228}_{-942} & 1.040 \pm 0.047 & 9212^{+394}_{-356} & 1.028 \pm 0.103 & 9569^{+922}_{-758} \\
24.0 & 7.599 \pm 0.161 & 14.901 \pm 0.056 & 14.518 \pm 0.058 & 14.178 \pm 0.027 & 1.105 \pm 0.104 & 8465^{+989}_{-788} & 1.075 \pm 0.045 & 9138^{+362}_{-333} & 1.032 \pm 0.100 & 9767^{+898}_{-744} \\
25.9 & 7.190 \pm 0.139 & 15.012 \pm 0.090 & 14.605 \pm 0.067 & 14.213 \pm 0.040 & 1.131 \pm 0.136 & 8349^{+1412}_{-1035} & 1.153 \pm 0.063 & 8757^{+500}_{-447} & 1.146 \pm 0.124 & 9170^{+975}_{-785} \\
26.2 & 7.264 \pm 0.142 & 15.022 \pm 0.076 & 14.609 \pm 0.066 & 14.210 \pm 0.040 & 1.131 \pm 0.123 & 8211^{+1192}_{-906} & 1.154 \pm 0.059 & 8646^{+445}_{-402} & 1.145 \pm 0.121 & 9093^{+942}_{-761} \\
28.1 & 7.220 \pm 0.138 & 15.049 \pm 0.076 & 14.614 \pm 0.057 & 14.223 \pm 0.037 & 1.159 \pm 0.111 & 7968^{+1039}_{-806} & 1.155 \pm 0.056 & 8644^{+424}_{-383} & 1.136 \pm 0.109 & 9166^{+844}_{-696} \\
29.7 & 5.986 \pm 0.177 & 15.148 \pm 0.073 & 14.643 \pm 0.054 & 14.170 \pm 0.012 & 1.456 \pm 0.125 & 7333^{+815}_{-655} & 1.505 \pm 0.059 & 7965^{+324}_{-263} & 1.535 \pm 0.106 & 8368^{+638}_{-485} \\
30.1 & 6.719 \pm 0.115 & 15.166 \pm 0.065 & 14.642 \pm 0.054 & 14.169 \pm 0.013 & 1.313 \pm 0.101 & 7189^{+709}_{-593} & 1.353 \pm 0.037 & 7861^{+249}_{-232} & 1.368 \pm 0.090 & 8370^{+649}_{-488} \\
31.2 & 6.312 \pm 0.111 & 15.240 \pm 0.080 & 14.658 \pm 0.058 & 14.181 \pm 0.018 & 1.439 \pm 0.122 & 6791^{+756}_{-613} & 1.449 \pm 0.048 & 7728^{+322}_{-275} & 1.452 \pm 0.104 & 8340^{+699}_{-521} \\
32.1 & 6.041 \pm 0.093 & 15.202 \pm 0.076 & 14.625 \pm 0.059 & 14.193 \pm 0.026 & 1.524 \pm 0.126 & 6813^{+734}_{-596} & 1.485 \pm 0.052 & 7924^{+317}_{-292} & 1.447 \pm 0.114 & 8723^{+710}_{-592} \\
33.9 & 6.093 \pm 0.096 & 15.317 \pm 0.070 & 14.696 \pm 0.056 & 14.131 \pm 0.033 & 1.508 \pm 0.114 & 6499^{+606}_{-508} & 1.613 \pm 0.054 & 7131^{+252}_{-233} & 1.670 \pm 0.101 & 7578^{+511}_{-440} \\
37.0 & 5.659 \pm 0.089 & 15.401 \pm 0.061 & 14.684 \pm 0.051 & 14.162 \pm 0.029 & 1.734 \pm 0.116 & 5938^{+444}_{-384} & 1.729 \pm 0.052 & 7045^{+214}_{-202} & 1.713 \pm 0.102 & 7908^{+498}_{-441} \\
39.8 & 5.553 \pm 0.075 & 15.516 \pm 0.067 & 14.757 \pm 0.054 & 14.195 \pm 0.028 & 1.749 \pm 0.125 & 5736^{+448}_{-391} & 1.766 \pm 0.049 & 6809^{+212}_{-197} & 1.776 \pm 0.098 & 7598^{+475}_{-420}
\enddata
\end{deluxetable*}

%% file: priors.tex
\begin{deluxetable*}{lLlllLl}
\tablecaption{Shock Cooling Parameters\label{tab:prior}}
\tablehead{\colhead{Parameter} & \colhead{Variable} & \colhead{Prior Shape} & \twocolhead{Prior Parameters\tablenotemark{a}} & \colhead{Best-fit Value\tablenotemark{b}} & \colhead{Units}}
\startdata
Shock velocity & v_\mathrm{s*} & Uniform & 0 & 3 & 1.0 \pm 0.2 & $10^{8.5}$ cm s$^{-1}$ \\
Envelope mass\tablenotemark{c} & M_\mathrm{env} & Uniform & 0 & 10 & 0.6 \pm 0.1 & $M_\sun$ \\
Ejecta mass $\times$ numerical factor\tablenotemark{d} & f_\rho M & Uniform & 3 & 100 & 60 \pm 30 & $M_\sun$ \\
Progenitor radius & R & Uniform & 0 & 100 & 14 \pm 2 & $10^{13}$ cm \\
Distance & d_L & Gaussian & 23.4 & 4.9 & 25 \pm 5 & Mpc \\
Extinction & E(B-V) & Gaussian & 0.104 & 0.0155 & 0.105^{+0.010}_{-0
.009} & mag \\
Explosion time & t_0 & Uniform & 59460 & 59464.6 & 59464.40 \pm 0.06 & MJD \\
Intrinsic scatter & \sigma & Half-Gaussian & 0 & 1 & 4.8 \pm 0.2 & Dimensionless \\
\enddata
\tablenotetext{a}{The ``Prior Parameters'' column lists the minimum and maximum for a uniform distribution, and the mean and standard deviation for a Gaussian distribution.}
\tablenotetext{b}{The ``Best-fit Value'' column is determined from the 16th, 50th, and 84th percentiles of the posterior distribution, i.e., $\mathrm{median} \pm 1\sigma$.}
\tablenotetext{c}{\cite{sapir_uv/optical_2017} define the envelope of an RSG as the region where $\delta \equiv \frac{R-r}{R} \ll 1$, where the density follows $\rho_0(\delta) = \frac{3 f_\rho M}{4 \pi R^3} \delta^n$ (we adopt $n=\frac{3}{2}$ for convective envelopes).}
\tablenotetext{d}{The ejecta mass and density profile do not have a strong effect on the light curve. Therefore this parameter is essentially unconstrained.}
\end{deluxetable*}

%% file: spectroscopy.tex
\begin{deluxetable*}{ccccccc}
\tabletypesize\scriptsize
\tablecaption{Log of Spectroscopic Observations\label{tab:spec}}
\tablecolumns{7}
\tablehead{\colhead{MJD} & \colhead{Telescope} & \colhead{Instrument} & \colhead{Phase} & \colhead{Exposure} & \colhead{Wavelength} & \colhead{Resolution} \\[-10pt]
\colhead{} & \colhead{} & \colhead{} & \colhead{(d)} & \colhead{Time (s)} & \colhead{Range (nm)} & \colhead{$\lambda/\Delta\lambda$}}
\startdata
59466.507 & FTN & FLOYDS & 2.1 & 600& 335--930 & 380 \\
59466.596 & FTS & FLOYDS & 2.2 & 600 & 335--930 & 250 \\
59467.643 & FTS & FLOYDS & 3.2 & 900 & 335--930 & 250 \\
59467.680 & SSO 2.3\,m & WiFeS & 3.3 & 3600 & 559--850 & 3000 \\
59468.484 & MMT & Binospec & 4.1 & 1440 & 383--920 & 650 \\
59468.769 & FTS & FLOYDS & 4.3 & 900 & 335--930 & 250  \\
59469.480 & MMT & Binospec & 5.1 & 1440 & 569--721 & 2900 \\
59469.606 & FTS & FLOYDS & 5.2 & 900 & 335--930 & 250  \\
59471.312 & NTT & EFOSC & 6.9 & 900 & 338--932 & 350 \\
59471.590 & FTS & FLOYDS & 7.1 & 900 & 335--930 & 250  \\
59471.652 & Keck II & KCWI & 7.2 & 60 & 350--560 & 1250 \\
59472.653 & FTS & FLOYDS & 8.2 & 1500 & 335--930 & 250 \\  
59474.485 & Keck II & NIRES & 10.0 & 1200 & 965--2460 & 2700 \\
59474.549 & FTN & FLOYDS & 10.1 & 900 & 335--930 & 380 \\
59475.734 & FTS & FLOYDS & 11.3 & 900 & 335--930 & 250 \\
59476.975 & SALT & RSS & 12.5 & 1495 & 350--930 & 1000 \\
59477.461 & MMT & MMIRS & 13.0 & 1200 & 940--1510 & 960 \\
59477.542 & FTN & FLOYDS & 13.1 & 900 & 335--930 & 380 \\
59478.484 & FTN & FLOYDS & 14.0 & 600 & 335--930 & 460 \\
59479.552 & FTN & FLOYDS & 15.1 & 900 & 335--930 & 380 \\
59479.969 & SALT & RSS & 15.5 & 1495 & 350--930 & 1000 \\
59480.641 & FTS & FLOYDS & 16.1 & 600 & 335--930 & 270 \\
59481.332 & NTT & EFOSC & 16.8 & 900 & 338--932 & 350 \\
59481.961 & SALT & RSS & 17.5 & 1495 & 350--930 & 1000 \\
59482.276 & NTT & SOFI & 17.8 & 960 & 950--2520 & 600 \\
59482.515 & FTN & FLOYDS & 18.0 & 900 & 335--930 & 380 \\
59483.490 & Shane & Kast & 19.0 & 630/600 & 300--1050 & 600 \\
59483.557 & FTN & FLOYDS & 19.0 & 600 & 335--930 & 460 \\
59486.549 & FTN & FLOYDS & 22.0 & 600 & 335--930 & 460 \\
59487.519 & FTN & FLOYDS & 23.0 & 900 & 335--930 & 380 \\
59488.550 & FTN & FLOYDS & 24.0 & 600 & 335--930 & 460 \\
59490.474 & FTN & FLOYDS & 25.9 & 900 & 335--930 & 380 \\
59490.721 & FTS & FLOYDS & 26.2 & 600 & 335--930 & 270 \\
59492.641 & FTS & FLOYDS & 28.1 & 2400 & 335--930 & 250 \\  
59493.371 & Bok & B\&C & 28.8 & 1000 & 601--716 & 3400 \\
59494.301 & Bok & B\&C & 29.7 & 1000 & 340--759 & 700 \\
59494.634 & FTS & FLOYDS & 30.1 & 600 & 335--930 & 270 \\
59495.760 & FTS & FLOYDS & 31.2 & 900 & 335--930 & 250 \\
59496.726 & FTS & FLOYDS & 32.1 & 600 & 335--930 & 270 \\
59498.522 & FTN & FLOYDS & 33.9 & 1500 & 335--930 & 380 \\  
59501.566 & FTN & FLOYDS & 37.0 & 900 & 335--930 & 380 \\
59504.475 & FTS & FLOYDS & 39.8 & 600 & 335--930 & 270 \\
59508.544 & FTN & FLOYDS & 43.9 & 900 & 335--930 & 380 \\
59509.457 & Keck I & HIRES & 44.8 & 3001 & 364--799 & 50,000 \\
59516.578 & FTS & FLOYDS & 51.9 & 600 & 335--930 & 270 \\
59516.668 & SSO 2.3\,m & WiFeS & 52.0 & 3600 & 350--900 & 3000 \\
59517.522 & FTS & FLOYDS & 52.8 & 900 & 335--930 & 250 \\
59517.598 & SSO 2.3\,m & WiFeS & 52.9 & 3600 & 350--900 & 3000 \\
59519.524 & FTN & FLOYDS & 54.8 & 600 & 335--930 & 460 \\
59524.458 & FTN & FLOYDS & 59.7 & 900 & 335--930 & 380 \\
59526.339 & Shane & Kast & 61.6 & 1703/1698 & 300--1050 & 600 \\
59533.355 & Shane & Kast & 68.6 & 630/600 & 300--1050 & 600 \\
59533.719 & FTS & FLOYDS & 68.9 & 900 & 335--930 & 250 \\
59543.294 & Shane & Kast & 78.4 & 630/600 & 300--1050 & 600 \\
59551.642 & FTS & FLOYDS & 86.7 & 900 & 335--930 & 250 \\
59560.422 & FTS & FLOYDS & 95.5 & 900 & 335--930 & 250 \\
59565.133 & SOAR & TripleSpec & 100.2 & 960 & 940--2460 & 3500 \\
59565.273 & MMT & MMIRS & 100.3 & 1440 & 940--1510 & 960 \\
59566.051 & SOAR & Goodman & 101.1 & 600 & 300--705 & 850 \\
59566.237 & MMT & MMIRS & 101.3 & 1440 & 940--1510 & 960 \\
59571.537 & FTS & FLOYDS & 106.5 & 900 & 335--930 & 250 \\
59580.564 & FTS & FLOYDS & 115.5 & 900 & 335--930 & 250 \\
59582.461 & FTS & FLOYDS & 117.4 & 1000 & 335--930 & 250 \\
59586.100 & Shane & Kast & 121.0 & 630/600 & 300--1050 & 600 \\
59593.488 & FTS & FLOYDS & 128.4 & 1000 & 335--930 & 250 \\
59605.103 & Shane & Kast & 139.9 & 1230/1200 & 300--1050 & 600 \\
59605.144 & ARC 3.5\,m & DIS & 139.9 & 1800 & 380--950 & 1000 \\
59611.128 & Shane & Kast & 145.9 & 930/900 & 300--1050 & 600 \\
\enddata
\tablecomments{Wavelength range and resolution are approximate. Exposure times for Kast spectra are for the blue and red halves, respectively. \explain{Replaced configuration column with exposure time, wavelength range, and resolution. Removed the cut-in headers.}}
\end{deluxetable*}